%% file: main.tex
\documentclass{jpp}
\usepackage{graphicx}
\usepackage[utf8]{inputenc}
\usepackage[T1]{fontenc}
\usepackage{amsmath}
\usepackage[dvipsnames]{xcolor}
\usepackage{pgfplots, tikz}
\usepackage{subfigure, dcolumn, bm, color, accents, hyperref, tikz ,booktabs}
\hypersetup{
    colorlinks   = true,
    citecolor    = blue,
    linkcolor    = blue}

\shorttitle{Mirror Confinement with Lenard-Bernstein Collisions}

\shortauthor{M. H. Rosen, W. Sengupta, I. Ochs, F. I. Parra, and G. W. Hammett}

\title{Enhanced Collisional Losses from a Magnetic Mirror Using the Lenard-Bernstein Collision Operator}

\author{Maxwell H. Rosen\aff{1}
    \corresp{\email{mhrosen@pppl.gov}},
    W. Sengupta\aff{1},
    I. Ochs\aff{1},
    F. I. Parra\aff{1,2}, \\
    \and G. W. Hammett\aff{2}}

\affiliation{\aff{1}Department of Astrophysical Sciences, Princeton University, Princeton, NJ 08540, USA
\aff{2} Princeton Plasma Physics Laboratory, Princeton, NJ 08540, USA}

\begin{document}

\maketitle

\begin{abstract}
    \input{abstract}
\end{abstract}

\input{introduction}
\input{methods}
\input{numerics}
\input{discussion}
\input{conclusion}

\section{Acknowledgements}
The authors thank Ammar Hakim, Manaure Francisquez, and Igor Kaganovich for their helpful discussions culminating in the ideas presented here. This work was supported by Princeton University and the U.S. Department of Energy under contract number DE-AC02-09CH11466. The United States Government retains a non-exclusive, paid-up, irrevocable, world-wide license to publish or reproduce the published form of this manuscript, or allow others to do so, for United States Government purposes. W.S. was supported by a grant from the Simons Foundation/SFARI (560651, AB) and DoE Grant No. DE-AC02-09CH11466. 

The code and data used to generate the figures in this publication are stored on a private GitHub repository. The source code is available upon request and consultation with the authors.

\input{appendix}

\bibliographystyle{jpp}
\bibliography{reference}

\end{document}

%% file: abstract.tex
Collisions are crucial in governing particle and energy transport in plasmas confined in a magnetic mirror trap. Modern gyrokinetic codes model transport in magnetic mirrors, but some utilize approximate model collision operators. This study focuses on a Pastukhov-style method of images calculation of particle and energy confinement times using a Lenard-Bernstein model collision operator. Prior work on parallel particle and energy balances used a different Fokker-Planck plasma collision operator. The method must be extended in non-trivial ways to study the Lenard-Bernstein operator. To assess the effectiveness of our approach, we compare our results with a modern finite element solver. Our findings reveal that the particle confinement time scales like $a \exp(a^2)$ using the Lenard-Bernstein operator, in contrast to the more accurate scaling that the Coulomb collision operator would yield $a^2 \exp(a^2)$, where $a^2$ is approximately proportional to the ambipolar potential. We propose that codes solving for collisional losses in magnetic mirrors utilizing the Lenard-Bernstein or Dougherty collision operator scale their collision frequency of any electrostatically confined species. This study illuminates the collision operator's intricate role in the Pastukhov-style method of images calculation of collisional confinement.

%% file: introduction.tex
\section{Introduction}
Magnetic mirrors, or adiabatic traps, present a compelling avenue for plasma confinement through the deflection of particles away from high-field regions. In recent years, the resurgence of interest in mirrors as a fusion concept has been led by groundbreaking experiments at the collisional Gas Dynamic Trap Experiment (GDT) at the Budker Institute in Novosibirsk, Russia, which achieved unprecedented transient electron temperatures of 900 eV, demonstrating the viability of mirrors in fusion endeavors \citep{Bagryansky2015}.

One of the most remarkable results from the GDT experiment is the stabilization of axisymmetric mirrors against the well-understood interchange instability \citep{post1966electrostatic}. Vortex confinement, for instance, has been shown to stabilize the $m=1$ flute interchange mode, and finite Larmor radius effects can stabilize $m\geq 2$ \citep{beklemishev2010vortex,bagryansky2011confinement, Beklemishev.2017,ryutov2011magneto, white2018centrifugal}. Moreover, recent advancements in superconducting magnet technology and electron cyclotron heating (ECH) have motivated new experiments to extend the results of GDT \citep{fowler2017new}. One such experiment is the Wisconsin high-field axisymmetric mirror experiment (WHAM) in Madison, Wisconsin \citep{egedal2022fusion, endrizzi2023physics}. Their new endeavor, utilizing high-temperature superconducting (HTS) REBCO tapes and neutral beam injection (NBI), will investigate the magnetohydrodynamic (MHD) and kinetic stability of the axisymmetric mirror, extended into the collisionless regime. With these new experimental techniques and MHD stability in sight, questions arise related to particle and energy confinement in these new stable axisymmetric mirror configurations.

Parallel losses play a critical role in the confinement of particles and energy, which occur when particles scatter due to collisions across the loss cone \citep{berk1988dissipative}.  \citet{pastukhov1974collisional} laid the foundation for calculating parallel losses in a magnetic mirror with the method of images approach that this study utilizes. Pastukhov's insight showed that the steady-state balance between a low-energy source, Fokker-Planck collision operator, and high-energy image sinks could be simplified by transforming the problem into an analogous Poisson equation, then solved using standard techniques from electromagnetism \citep{jackson1999classical}. Building on Pastukhov's insight, \citet{najmabadi1984collisional} extended the analysis by simplifying Pastukhov's variable transformations, reducing the number of approximations made, and including a higher-order correction, yielding a more refined solution. Recent advancements have been made by \citet{ochs2023confinement}, who included relativistic effects to Najmabadi's approach.

Although parallel dynamics are often the fastest time-scale phenomenon in magnetized plasmas, many researchers are interested in studying the next-order perpendicular transport due to micro-stability and turbulence, which find their best answers in computer simulations that include collisions. With the recent investigations using the Gkeyll code to study high-field magnetic mirrors utilizing the Dougherty collision operator \citep{francisquez2023towards}, it is essential to understand the effect of this approximate collision operator on parallel collisional losses before it is extended to study perpendicular transport.

To understand the key details and trade-offs between different collision operators, we must first study a few important approximations in this context. A comprehensive description of two-particle collisions within the framework of a Fokker-Planck operator involves the so-called `Rosenbluth potentials' \citep{rosenbluth1957fokker}. This full-fledged operator, while widely studied and implemented, poses computational and analytical challenges \citep{taitano2015mass}. In some cases, approximations become a pragmatic necessity to render the problem computationally tractable. In this context, one widely used approximate collision operator is the Lenard-Bernstein (LBO) / Dougherty operator: a simple operator that captures the advection and diffusion responses from small-angle two-body collisions \citep{lenard1958plasma,dougherty1964model}. The difference between the LBO and Dougherty collision operators is that the LBO has zero parallel streaming fluid flow velocity, while the Dougherty operator includes the parallel fluid flow velocity in calculating the drag, useful for cases where momentum conservation is important.

Novel work was performed using the Gkeyll code in projecting the Dougherty operator onto a discontinuous Galerkin framework with enhanced multi-species collisions for gyrokinetic and Vlasov-Maxwell simulations \citep{hakim2020conservative,francisquez2020conservative,francisquez2022improved}. Owing to the Dougherty operator having eigenfunctions of Hermite polynomials, the GX code uses this simple collision operator \citep{mandell2022gx}. The GENE-X code can simulate x-point geometry tokamak configurations, with their most rigorous collision operator being Dougherty \citep{ulbl2022implementation, ulbl2023influence}. Other examples of plasma kinetic/gyrokinetic codes that have implemented such collision operators (at least as an option) include: \citet{Ye2024}, \citet{Celebre2023}, \citet{Hoffmann_Frei_Ricci_2023}, \citet{Frei_Hoffmann_Ricci_2022}, \citet{Perone2020}, \citet{LOUREIRO2016}, \citet{GRANDGIRARD201635}, \citet{Pezzi2016}, \citet{Parker_Dellar_2015}, and \citet{Hatch2013}.
An LBO / Dougherty collision operator with the appropriate definitions satisfies many properties of a good collision operator, such as conservation of density, momentum, and energy \citep{francisquez2020conservative}. The most significant defect in this model is that the collision frequency is independent of velocity, leading to inaccurate results in the tail of the distribution function. Furthermore, the operator's isotropic diffusion coefficient makes no distinction between pitch-angle scattering and energy diffusion \citep{hirshman1976approximate}. Some of the shortcomings of the Lenard-Bernstein/Dougherty operator are described in \citet{Knyazev2023}.

In this study, we build upon the method of images approach developed by \citet{najmabadi1984collisional}, but with a focus on the LBO, since we consider a system with zero parallel fluid flow. Since the ambipolar potential of magnetic mirrors shifts the loss cone towards higher-energy particles, we must investigate the validity of the LBO's collisionality approximation at high energies. Following the method of images approach from \citet{najmabadi1984collisional}, results show that the particle confinement time scales like $a \exp(a^2)$ using the LBO, in contrast to the scaling $a^2 \exp(a^2)$ that more accurate Coulomb collision operator yields, where $a^2$ is proportional to the ambipolar potential. In addition, the average energy of lost particles is also modified. The error between the average energy of lost particles and our numerical solver is comparable to the study of \citet{najmabadi1984collisional}. It is critical that a code utilizing the LBO or Dougherty collision operator matches particle loss rates compared to an experiment to predict the correct ambipolar potential. Suggestions are made to reduce the collision frequency to match particle loss rates compared to the results in \citet{najmabadi1984collisional}.

It is of interest to mention and discuss an alternative to the method of images approach, as the historical journey toward accurate ambipolar estimates encompasses diverse approaches. \citet{chernin1978ion} introduced an alternative derivation that approximates the loss cone as a square in velocity space, yielding insights using a linearized Fokker-Planck equation and the associated variational principle. \citet{cohen1978collisional} showed that \citet{chernin1978ion} techniques lacked robustness compared to Pastukhov's technique, and higher-order approximations were needed. Subsequent work in \citet{catto1981collisional}, followed by \citet{catto1985particle} and \citet{fyfe1981finite}, refined this approach by eliminating the square loss cone approximation and extending the approach to higher order. Catto's studies circumvent the need for an accurate solution for the distribution function by transforming the problem into parallel and perpendicular coordinates to the loss cone and then employing variational techniques to yield improved confinement times. \citet{khudik1997longitudinal} demonstrated the comparable accuracy of fourth-order extensions of Catto's methods to Najmabadi's approach. Furthermore, these variational methods generalize to arbitrary mirror ratios and, therefore, would be more suitable for application to toroidal confinement devices, which have order unity mirror ratios. Extending these models to the LBO would not be trivial because these methods are finely tuned to the Coulomb operator. Although these methods have many good properties, we are ultimately interested in large mirror ratios, and the flexibility of variational techniques is unnecessary.

Subsequent sections will use the method of images approach to explore how the LBO collision operator changes the parallel losses of a magnetic mirror. In section \ref{sec: methods of images}, the problem is presented and solved systematically for the confinement time and energy loss rate. A correction factor is evaluated numerically in section \ref{sec: correction factors}. Results are compared to the numerical code presented in section \ref{sec: numerics}. This approach's validity and applicability to mirror simulation codes are discussed in section \ref{sec: discussion}. Suggestions are made for modifying the collision frequency to obtain the appropriate ambipolar potential. Finally, we conclude in section \ref{sec: conclusion}.

%% file: methods.tex
\section{Method of Images Solution}\label{sec: methods of images}
Our solution follows closely with the methods of \citet{najmabadi1984collisional} with some key differences. In a single-particle picture of a magnetic mirror, the parallel dynamics of particle losses depend on whether they possess sufficient energy to overcome the confining forces. These forces arise from gradients in the magnetic field magnitude $(\vec\nabla B)$ and an electrostatic ambipolar potential ($z_s e\phi$), where $e$ is the elementary charge and $z_s$ is the charge number of species $s$. When particles have enough energy to overcome the $\vec\nabla B$ forces in the absence of an ambipolar potential, they reside within a specific region in phase space known as the ``loss cone.'' The introduction of an ambipolar potential alters the minimum energy required for escape, thereby transforming the loss cone into a ``loss hyperboloid.'' The loss hyperboloid can be expressed as
\begin{align}
    1-\mu^2 = \frac{1}{R} \left( 1 - \frac{v_0^2}{v^2} \right). \label{eqn: loss cone}
\end{align}
\begin{figure}
\centering
\begin{tikzpicture}
    \centering
  \begin{axis}[grid=none, minor grid style={dashed,black!50}, major grid style={black!80}, ylabel={$v_\perp$}, xlabel={$v_{||}$}, ymin=-0.00, ymax=3.5, xmin=-7, xmax=7, trim axis left, trim axis right, xtick=\empty, ytick=\empty, width=1.8*\axisdefaultheight, height=\axisdefaultheight*0.9, extra x ticks = {-2, 0, 2}, extra x tick labels = {$-\sqrt{2 z_s e \phi / m_s}$, $0$, $\sqrt{2 z_s e \phi / m_s}$}]
    \def\m{40} 
    \def\b{4} 
    \def \sinkMin{3}
    \def \yMax{7}
    \addplot[domain=2:7, samples=100, RoyalBlue] {sqrt((x^2 - \b)/\m)};
    \addplot[domain=-7:-2, samples=100, RoyalBlue] {sqrt((x^2 - \b)/\m)};
    \addplot[domain=0:7, samples=100, BrickRed, dashed] {x / sqrt(\m)};
    \addplot[domain=-7:0, samples=100, BrickRed, dashed] {-x / sqrt(\m)};
    \draw[ForestGreen, dotted, line width=3pt, dash pattern=on 1pt off 2pt] (axis cs: (\sinkMin, 0) -- (axis cs: \yMax, 0) node[midway, above, black] {Sinks};
    \draw[ForestGreen, dotted, line width=2pt, dash pattern=on 1pt off 2pt] (axis cs: (-\sinkMin, 0) -- (axis cs: -\yMax, 0) node[midway, above, black] {Sinks};
    \node[align=center, anchor=west] at (axis cs: 5.6, 0.5) {Loss \\ region};
    \node[align=center, anchor=east] at (axis cs: -5.6, 0.5) {Loss \\ region};
    \addplot [only marks, mark=*] coordinates {(0,0)} node[pin=90:Source] {};
    \node[align=center, anchor=center] at (axis cs: 0,  2) {Confinement \\ region};
  \end{axis}
\end{tikzpicture}
\caption{The loss boundary in velocity space for electrostatically confined particles in a magnetic mirror field. Imposed on the figure is a model depiction of the low energy source. The dotted green line represents the sinks used to solve for the distribution function. The blue line is the loss hyperboloid described in equation \eqref{eqn: loss cone}, and the red dashed line is the loss cone without an ambipolar potential.}
\label{fig: loss cone}
\end{figure}

Here, $\mu = \cos \theta = v_{||} / v$ is the cosine of the pitch angle, $v$ is the total velocity $|\vec{v}|$, $v_{||} = \left(\vec{v} \cdot \vec{B} \right) / B$ is the component of the velocity parallel to the magnetic field, $R$ is the ratio of the maximum magnetic field to the minimum magnetic field $B_{\text{max}}/B_{\text{min}}$, and $v_0$ is the loss velocity corresponding to the ambipolar potential ($m_s v_0^2 /2 = z_s e \phi$), where $m_s$ is mass. The goal is to recreate this boundary, where the distribution function is null, with image sources and sinks. A model of the loss hyperboloid, source, and image sinks in the problem is shown in figure \ref{fig: loss cone}. A source is placed at a low velocity, in a steady state balance with the collision operator, while sinks are placed in the loss region and $f_s$ is extended. With these sinks, the equation for $f_s$ near the loss hyperboloid is
\begin{equation}
\mathcal{L}(f_s) + Q(v,\mu) = 0. \label{eqn: General equation L+Q=0}
\end{equation}
Here, $\mathcal{L}(f_s)$ is a collision operator which acts on the distribution function $f_s$ of species $s$, and $Q(v,\mu)$ is an image sink of particles placed inside the loss region, where $f_s$ is extended. The sink $Q$ will be chosen to make $f_s = 0$ at the loss hyperboloid's vertex defined by equation \eqref{eqn: loss cone} as well as have the contour of $f_s = 0$ match the radii of curvature of the loss hyperboloid at the vertex. Thus the boundary conditions on $f_s$ are defined as

\begin{align}
    &f_s({v},\mu)|_{v={v}_0,\mu=\pm 1} = 0, &\frac{\partial_{{v}} f_s}{\partial_\mu f_s}\bigg\rvert_{{v}=v_0,\mu=\pm 1} = \frac{1}{R {v}_0}. \label{eqn: BC distribution function}\
\end{align}
For small $v$, we assume that $f_s$ is a Maxwellian,
\begin{align}
    f_s(v,\mu) |_{v \rightarrow 0} \rightarrow \frac{n_s}{\pi^{3/2} v_{th,s}^2} \exp(-\frac{v^2}{v_{th,s}^2}), \label{eqn: boundary condition small velocity maxwellian}
\end{align}
where $n_s$ is number density, $v_{th,s} = \sqrt{2T_s / m_s}$, $T_s$ is temperature in units of energy, and $v$ is the total magnitude of the velocity $|\vec{v}|$. Assuming a square well approximation for the magnetic field and considering that the maximum magnetic field occurs at the mirror throat, the LBO has the form \citep{lenard1958plasma}
\begin{equation}
    \mathcal{L}(f_s) = \nu_{sLBO}\frac{\partial}{\partial \vec{v}} \cdot \left( \vec{v} f_s + \frac{v_{th,s}^2}{2} \frac{\partial f_s}{\partial \vec{v}} \right),
    \label{eqn:LBO}
\end{equation}
where $\nu_{sLBO}$ is the collision frequency used in the LBO, $\vec{v}$ is a velocity vector, and $v_{th,s}$ is the thermal velocity. Generality is left in defining the collision frequency $\nu_{sLBO}$ because it may be chosen to match certain important physical quantities and rates in its specific implementation \citep{francisquez2022improved}. For instance, one may choose the collision frequency for the LBO to match thermal equilibration rates, Braginskii heat fluxes, or the Spitzer resistivity. Later in this work, we will investigate choosing the LBO's collision frequency to match ambipolar collisional losses from a magnetic mirror field. To simplify the analysis, we adopt the following normalizations and definitions:
\begin{align}
    \bar{v} &= \frac{v}{v_{th,s}}; & F_s &= \frac{v_{th,s}^3 f_s}{n_s}.
\end{align}
Our normalization $\bar{v}$ is what \citet{pastukhov1974collisional} and others refer to as $x$. We restate the LBO in normalized spherical coordinates to improve clarity,

\begin{equation}
    \mathcal{L}(F_s) =  \frac{1}{\bar{v}^2} \frac{\partial}{\partial \bar{v}} \bar{v}^3 \left( F_s + \frac{1}{2\bar{v}} \frac{\partial F_s}{\partial \bar{v}}\right) + \frac{Z_s}{2 \bar{v}^2}  \frac{\partial}{\partial \mu} \left( 1 - \mu^2 \right) \frac{\partial F_s}{\partial \mu} \label{eqn: Dougherty collision operator}.
\end{equation}
We introduce a factor $Z_s$ into the diffusion term to correct for the LBO's approximation of treating the pitch-angle scattering and energy diffusion terms equally. Although for the LBO $Z_s=1$, an opportunity is left for future work to correct this defect by modifying the coefficient $Z_s$. The factor of 2 in the pitch-angle scattering in equation \eqref{eqn: Dougherty collision operator} comes from the $1/2$ in the $v_{th,s}^2$ term in equation \eqref{eqn:LBO}. An essential distinction between equation \eqref{eqn: Dougherty collision operator} and the collision operators proposed by \citet{pastukhov1974collisional} and \citet{najmabadi1984collisional} lies in the factors of $\bar{v}$. The collision operators in their work retain the velocity dependence within the Rosenbluth potentials. Below are the collision operators from  \citet{pastukhov1974collisional} and \citet{najmabadi1984collisional},

\begin{align}
        \mathcal{L}_{\text{Najmabadi}}(F_s) &= \frac{1}{\bar{v}^2} \frac{\partial}{\partial \bar{v}} \left( F_s + \frac{1}{2 \bar{v}} \frac{\partial F_s}{\partial \bar{v}}\right) + \frac{1}{\bar{v}^3} \left( Z_{s,N} - \frac{1}{4 \bar{v}^2}\right)\frac{\partial}{\partial \mu} \left( 1 - \mu^2 \right) \frac{\partial F_s}{\partial \mu}, \label{eqn: Collision operator najmabadi} \\
        \mathcal{L}_{\text{Pastukhov}}(F_s) &= \frac{1}{\bar{v}^2} \frac{\partial}{\partial \bar{v}} \left( F_s + \frac{1}{2 \bar{v}} \frac{\partial F_s}{\partial \bar{v}}\right) + \frac{1}{\bar{v}^3} \frac{\partial}{\partial \mu} \left( 1 - \mu^2 \right) \frac{\partial F_s}{\partial \mu}. \label{eqn: Collision operator pastukhov}
\end{align}
where $Z_{s,N}$ is the $Z$ that is used in \citet{najmabadi1984collisional} and $N$ stands for \citet{najmabadi1984collisional}, detailed in appendix \ref{sec: Najmabadi correction}. To compare, the collision operator used in \citet{pastukhov1974collisional} treats the factor in equation \eqref{eqn: Collision operator najmabadi} $\left(Z_{s,N} - 1/(4 \bar{v}^2)\right)$ as one, although \citet{cohen1986interchange} addresses this limitation in treating the multi-species collisions. Comparing equation \eqref{eqn: Collision operator najmabadi} and equation \eqref{eqn: Dougherty collision operator}, we see that the drag and parallel diffusion are missing the $1/\bar{v}^3$ scaling of the more accurate collision operators.  However, the pitch angle scattering term is not as bad, scaling like $1/\bar{v}^2$ for the LBO and like $1/\bar{v}^3$ for the more accurate operators.

To simplify the solution, we define a general form for the image problem. We impose that the image sinks start at velocity $a$, are placed solely outside the loss hyperboloid ($a > \bar{v}_0$ where $\bar{v}_0^2 = z_s e \phi / T_s$), and are isolated to lie along $\mu = \pm 1$ to preserve the symmetry of the problem. As described in figure \ref{fig: loss cone},
\begin{equation}
    Q(\bar{v},\mu) = -\frac{\delta(1-\mu^2)}{4\pi} H(\bar{v}-a) q(\bar{v}), \label{eqn:source def}
\end{equation}
where $H()$ is the Heaviside step function and $\delta()$ is the Dirac delta function. 

In \citet{pastukhov1974collisional}, they assume a form for the sinks $q(\bar{v}) = q_0 \exp(-\bar{v}^2)$, but the later work by \citet{najmabadi1984collisional} shows that defining $q(\bar{v}) = q_0 \exp(-\bar{v}^2) (Z_a - 1/4 \bar{v}^2)/ \bar{v}^3$ makes the resultant equations simpler to solve with fewer approximations. \citet{najmabadi1984collisional} finds the form of $q(\bar{v})$ by leaving its functional form free during the problem setup, then choosing a specific form at a later stage to facilitate a solution. Thus, we choose to leave the form of $q(\bar{v})$ arbitrary in equation \eqref{eqn:source def} and will define its form at a later stage.

Without approximation, we can use equation \eqref{eqn: Dougherty collision operator} to rewrite equation \eqref{eqn: General equation L+Q=0} in the following way.
\begin{equation}
    \frac{2 e^{\bar{v}^2}}{\bar{v}}  \frac{\partial}{\partial e^{\bar{v}^2}} \bar{v}^3 \frac{\partial}{\partial e^{\bar{v}^2}} \left( e^{\bar{v}^2} F_s \right) + \frac{Z_s}{2 \bar{v}^2}  \frac{\partial}{\partial \mu} \left( 1 - \mu^2 \right) \frac{\partial F_s}{\partial \mu}  + Q(\bar{v},\mu) = 0.
\end{equation}
The inverse chain rule is used to absorb factors of $\bar{v}$ into the derivatives, and we also employ the identity  $ F_s + (1/ 2 \bar{v}) \partial_{\bar{v}} F_s = (1/ 2 \bar{v} ) \exp{(-\bar{v}^2)}\partial_{\bar{v}} (\exp{(\bar{v}^2)} F_s)$. Let us define here the variable transformation $z(\bar{v})=\exp{(\bar{v}^2)}$. We now make a critical approximation: large changes in $z$ cause only small changes in $\bar{v}^2 = \ln(z)$, making the derivatives of powers of $\bar{v}$ small. This justifies moving the $\bar{v}^3$ outside the derivative. In more detail, the approximation we make says
\begin{equation}
    3 \bar{v}^2 F_s , \bar{v} \frac{\partial F_s}{\partial \bar{v}} \ll \bar{v}^3 \frac{\partial F_s}{\partial \bar{v}} , \bar{v}^3 \frac{\partial}{\partial \bar{v}} \left( \frac{1}{2 \bar{v}} \frac{\partial F_s}{\partial \bar{v}} \right). \label{eqn:xcubed approximation}
\end{equation}
This is valid since we are only interested in $F_s$ near the loss cone at large velocities, where $\partial F_s / \partial \bar{v}$ is large and $\bar{v} \ll \bar{v}^3$. To simplify further, we define $g(\bar{v},\mu) = \pi^{3/2} \exp(\bar{v}^2) F_s(\bar{v},\mu)$ to factor out the Maxwellian component of the solution. This leads to the new form
\begin{align}
\left( \frac{\partial^2}{\partial z^2}  + \frac{Z_s}{4 z^2 \ln(z)^2}  \frac{\partial}{\partial \mu} \left( 1 - \mu^2 \right) \frac{\partial}{\partial \mu} \right) g_s(\bar{v},\mu) + \frac{\pi^{3/2}}{2 z \ln(z)} Q(\bar{v},\mu) = 0.
\end{align}

We aim to make the operator on $g(\bar{v},\mu)$ resemble a cylindrical Laplacian to map this problem to a Poisson problem. For this reason, we must also perform a transformation on the pitch angle scattering component. Consider a general variable transform of the form $\rho(\bar{v},\mu) = h(\bar{v},\mu) \sqrt{1-\mu^2}$. Assuming a large mirror ratio $R \gg 1$, we may approximate that near the loss cone $\mu \approx \pm 1$. Thus, computing $\partial \rho/ \partial \mu$, we neglect $\partial h / \partial \mu$, which would be multiplied by a term $(1-\mu^2)$, which is small. From this, we find $\partial_\mu \approx -(\mu h^2/\rho )\partial_\rho$ and $(1-\mu^2) = \rho^2/h^2$. Under this transformation $(1-\mu^2)\partial_\mu$ becomes $- \mu \rho \partial_\rho$ without any factors of $h(\bar{v}, \mu)$. 

Part of the elegance of this method is that we are not limited by the form of $h(\bar{v},\mu)$, as long as $z$ is only a function of $\bar{v}$. For the purpose of an argument, consider an arbitrary transformation $z(\bar{v}, \mu)$.  If $\partial_\mu z$ were non-zero, we would have to use the chain rule and compute $\partial_\mu \rho(z,\mu) = \partial_z \rho \, \partial_\mu z+ \partial_\mu \rho$, thus complicating the procedure. In \citet{najmabadi1984collisional} and here, our variable transformation $z = \exp(\bar{v}^2)$ means that $ \partial_z \rho \, \partial_\mu z = 0$, making this variable transformation simpler. This insight is perhaps the most pivotal innovation in \citet{najmabadi1984collisional} compared to the variable transformations presented in \citet{pastukhov1974collisional}. \citet{pastukhov1974collisional} uses a variable transformation $z = \exp(\bar{v}^2) \mu / \sqrt{2 \bar{v}^2}$ and $\rho = \exp(\bar{v}^2) \sqrt{1 - \mu^2}$ where both variable transformations are functions of $\bar{v}$ and $\mu$. When ultimately satisfying boundary conditions, this leads \citet{pastukhov1974collisional} to do ``a number of straightforward but rather cumbersome algebraic transformations.'' The complications arise due to the prefactors in front of the logarithm in their equation (17) having a factor of $\bar{v}/ \mu$, which comes from the $\mu / \bar{v}$ in their variable transformation for $z$. In contrast, \citet{najmabadi1984collisional} uses the variable transformations $z = \exp(\bar{v}^2)$ and $\rho = \sqrt{ 2 \bar{v}^2 / (Z_a - 1/4 \bar{v}^2)} \exp(\bar{v}^2) \tan \theta$. Notice that, as we have pointed out above, the variable transformation in $z$ does not depend on $\mu$. The equivalent solution is \citet{najmabadi1984collisional} equation (23), which is divided by a Maxwellian compared to \citet{pastukhov1974collisional} equation (17). \citet{najmabadi1984collisional} equation (23) has a mere constant before the logarithm, making satisfying boundary conditions much simpler.

By setting $h(\bar{v},\mu)$ to cancel any factor in front of the pitch-angle scattering derivatives, it can be shown that the appropriate variable transformation is
\begin{align}
    \rho(\bar{v},\mu) &= \frac{2 z \ln(z)}{\sqrt{Z_s}} \frac{\sqrt{1 - \mu^2}}{\mu} = \frac{2 z \ln(z)}{\sqrt{Z_s}} \tan\theta. \label{eqn: rho transformation}
\end{align}
As a result, the problem is in the form of a cylindrical Poisson equation,
\begin{align}
    \frac{1}{\rho} \frac{\partial}{\partial \rho} \left(\rho \frac{\partial}{\partial \rho} g_s \right) + \frac{\partial^2 g_s}{\partial^2 z} = \frac{\pi^{3/2}}{2 z \ln(z)} \frac{\delta(1-\mu^2)}{4\pi} q (\bar{v}) H(\bar{v}-a). \label{eqn: Poisson distribution form}
\end{align}
We can now set the free function $q(\bar{v})$ such that the equation takes an ad-hoc, easy-to-solve form,
\begin{align}
     \frac{1}{\rho} \frac{\partial}{\partial \rho} \left(\rho \frac{\partial}{\partial \rho} g_s \right) + \frac{\partial^2 g_s}{\partial^2 z}= \frac{\delta(\rho)}{2 \pi \rho} 4 \pi q_0 H(z - z_a). \label{eqn: Poisson source form}
\end{align}
On the right-hand side of equation \eqref{eqn: Poisson source form}, $q_0$ is equivalent to a constant linear charge density and $z_a = \exp(a^2)$. Although we choose $q_0$ to be a constant for the sake of having an easy-to-solve problem, it does mean we are sacrificing some detail in the ability to match equation \eqref{eqn: loss cone} perfectly; however, exact matching is not necessary because the distribution function, as well as losses, decay at higher energies and the majority of particles are lost near the tip of the loss hyperboloid so that is the region we are most interested in matching. By matching equations \eqref{eqn: Poisson distribution form} and \eqref{eqn: Poisson source form}, we show in Appendix \ref{sec: Appendix 1} that the appropriate connection between $q(\bar{v})$ and $q_0$ is
\begin{align}
    q(\bar{v}) = q_0 \cdot \frac{8}{\sqrt{\pi}} \frac{\mu^2 Z_s}{z \ln(z)}. \label{eqn: q form}
\end{align}
With the new coordinates $\rho$ and $z$, we transform boundary condition \eqref{eqn: boundary condition small velocity maxwellian}. In addition, the lower limit of $z=1$ for $\bar{v}=0$ is extended to $z=0$ since $z \equiv \exp(\bar{v}^2) \gg 1$ or set $z^\prime = \exp(\bar{v}^2) - 1 = z(1 + O(\exp(-\bar{v}^2)))$.
\begin{align}
    g_s(\rho,z) |_{z=0} = 1 \label{eqn: boundary condition transformed g z=0}
\end{align}

Equation \eqref{eqn: Poisson source form} with boundary condition \eqref{eqn: boundary condition transformed g z=0} is solved in \citet{jackson1999classical}. The equivalent problem in electricity and magnetism terms is having a conducting boundary condition on the $z=0$ plane held at potential $g_s=1$ and placing a wire of constant linear charge density $q_0$ on the $z$-axis, suspended above the $z=0$ plane by height $z=z_a$. The standard method of solving this problem is to use the method of images to match the conducting boundary condition. To outline this approach, we use the method of images for the equivalent Poisson problem after previously using the method of image approach to place image sources and sinks in figure \ref{fig: loss cone}. This layering of the method of images is why this approach is referred to as the method of image solution for determining loss rates from a magnetic mirror. Only requiring unmapping the equation through stated variable transformations, we find the appropriate distribution function for a magnetic mirror as

\begin{align}
g_s(\rho,z) = 1 - q_0 \ln\left(\frac{z_a + z + \sqrt{\rho^2 + \left(z_a + z\right)^2}}{z_a - z + \sqrt{\rho^2 + \left(z_a - z\right)^2}}\right).\label{eqn: distg}
\end{align}

Mapping this problem back to the magnetic mirror and using that $z \gg 1$ near the loss hyperboloid, we can apply the boundary conditions assigned to this problem in equation \eqref{eqn: BC distribution function}. Interestingly, this is the only part of the calculation in which information about the magnetic field is used in the solution. We show in Appendix \ref{sec: BC Limits} that for a general variable transformation $z = \exp(\bar{v}^2)$, $\rho = \bar\rho(\bar{v}) z \tan \theta$, where $h(\bar{v}, \mu) = \bar\rho(\bar{v}) z / \mu$ mentioned earlier, we get
\begin{align}
    q_0 &= \left( \ln \left( \frac{w+1}{w-1}\right)\right)^{-1} \label{eqn: q0 bc matching}\\
    w^2 &=  1 + \frac{\bar\rho(\bar{v}_0)^2}{2 R \bar{v}_0^2}  = 1 + \frac{2 \bar{v}_0^2}{ Z_s R} \label{eqn: derivative bc matching} 
\end{align}
where $w = \exp (a^2 - \bar{v}_0^2)$. This can be inverted to give $a^2 = \bar{v}^2_0 + \ln(w)$, which determines the edge of the sink region, $a$, in terms of the tip of the loss region $v_0$.

In \citet{najmabadi1984collisional}, they adjust the strength of the sinks by examining the difference between the flux of the true loss cone, equation \eqref{eqn: loss cone}, and the approximate loss cone, where equation \eqref{eqn: distg} equals zero. They account for this in equations 41a and 41b. Their instructions are to evaluate \eqref{eqn: distg} along the loss cone one $z_se \phi / T_s$ above the tip of the loss cone in the limit where $R \gg 1$. This means $\bar{v}^2 = v_0^2+1$, $\mu = 1$, and $a \approx v_0$, which leads to $z = \exp(v_0^2 + 1)$ and $\rho = 0$, meaning $g(0,v_0^2+1) \approx 1 - q_0 \ln((e+1)/(e-1)) = 1 - 0.77 q_0$. Thus, we modify $q_0$ by subtracting $0.77$, leading to a corrected definition of $q_0$. \citet{najmabadi1984collisional} determines this to be $0.84 q_0$, but we have not replicated the calculation used to determine their equation (41b).
\begin{align}
    q_0 &= \left( \ln \left( \frac{w+1}{w-1}\right)- 0.77\right)^{-1}. \label{eqn: q0 bc matching with 0.77}
\end{align}

To calculate this system's confinement time and energy loss rate, we integrate the image sinks over all velocity space, considering the symmetry in gyro- and pitch-angle.
\begin{align}
\frac{1}{n_s \nu_{sLBO}}  \frac{dn_s}{dt} &= 2 \pi \int_0^\infty \bar{v}^2 d\bar{v} \int_{-1}^1 d\mu Q(\bar{v},\mu) \label{eqn: dndt definition}\\ 
\frac{3}{2}\frac{1}{\nu_{sLBO}}\frac{1}{n_s T_s} \frac{d n_s T_s}{dt} &= 2 \pi \int_0^\infty \bar{v}^4 d\bar{v} \int_{-1}^1 d\mu Q(\bar{v},\mu)
\end{align}
Although this appears to be a straightforward integral, there is a subtlety to handling it worth mentioning. Equation \eqref{eqn: dndt definition} appears as if it is integrating over half of each delta function on both sides, but it is, in fact, integrating two whole peaks of the delta function across all of velocity space, so $\int_{-1}^1 \delta(1-\mu^2) d\mu = 1$. Once accounting for this intricacy, we arrive at the following forms for confinement time and energy loss rate.
\begin{align}
    \frac{1}{\tau_{c}} = \frac{1}{n_s} \frac{dn_s}{dt} &= -\nu_{sLBO} 2 Z_s \frac{\mathrm{Erfc}(a)}{\ln \left( \frac{w+1}{w-1}\right)-0.77} ,\label{eqn:elegant loss rates rosen} \\
    \frac{1}{E_s}\frac{dE_s}{dt}= \frac{1}{n_s T_s} \frac{d (n_s T_s)}{dt} &= -\nu_{sLBO} \frac{4}{3 \sqrt{\pi}} Z_s \frac{ e^{-a^2}a + \frac{\sqrt{\pi}}{2} \mathrm{Erfc}(a)}{\ln \left( \frac{w+1}{w-1}\right)-0.77}.\label{eqn: energy loss rate}
\end{align}
Here, $\tau_c$ is the confinement time, $E_s = 3/2 \; n_sT_s$ is total energy of the system, $\mathrm{Erfc}()$ is the complementary error function, $w = \sqrt{1 + 2 z_s e \phi / \left( T_s Z_s R \right)}$, and $a = \sqrt{ z_s e\phi / T_s + \ln{(w)}}$. 

We proceed to calculate the energy of the particle, but due to collisions, $z_s e \phi / T_s \sim (1/2) \ln (m_i / m_e) \gg 1$. Furthermore, for $R \gg z_s e \phi / T_s Z_s$, $\ln(w)$ is small, giving $a \simeq \sqrt{z_s e \phi / T_s} \gg 1$. The asymptotic expansion for a large argument of the complementary error function is $\mathrm{Erfc}(x) \approx \exp(-x^2) / (\sqrt{\pi} x) \times (1 - 1/2x^2 + 3/4x^4 + \mathcal{O}(1/x^6) )$, so $\tau_c$ scales as $a \exp(a^2)$.  The average energy of lost particles is found by evaluating $E_{s,\text{loss}} = (dE_s/dt)/(dn_s/dt)$.

\begin{align}
    \frac{E_{s,\text{loss}}}{T_s} &= \frac{1}{2} \left( 1 + \frac{2 a e^{-a^2}}{\sqrt{\pi}\mathrm{Erfc}(a)}\right)\label{eqn: average loss energy}\\
    &\approx  a^2 + 1 - \frac{1}{2 a^2} + \mathcal{O}(1/a^4)\label{eqn: average loss energy asymptotic}
\end{align}

It is important to note that equation \eqref{eqn:elegant loss rates rosen} and subsequent definitions are presented in their un-normalized form. During the un-normalization process, all quantities are stated in terms of the temperature $T_s$ of the species. This avoids confusion when using a different normalization procedure for the thermal velocity.

%% file: numerics.tex
\section{Numerical simulations and corrections to equation \eqref{eqn: q0 bc matching}}\label{sec: correction factors}
\label{sec: numerics}

\begin{figure}
    \centering
    \includegraphics[width=0.9\linewidth]{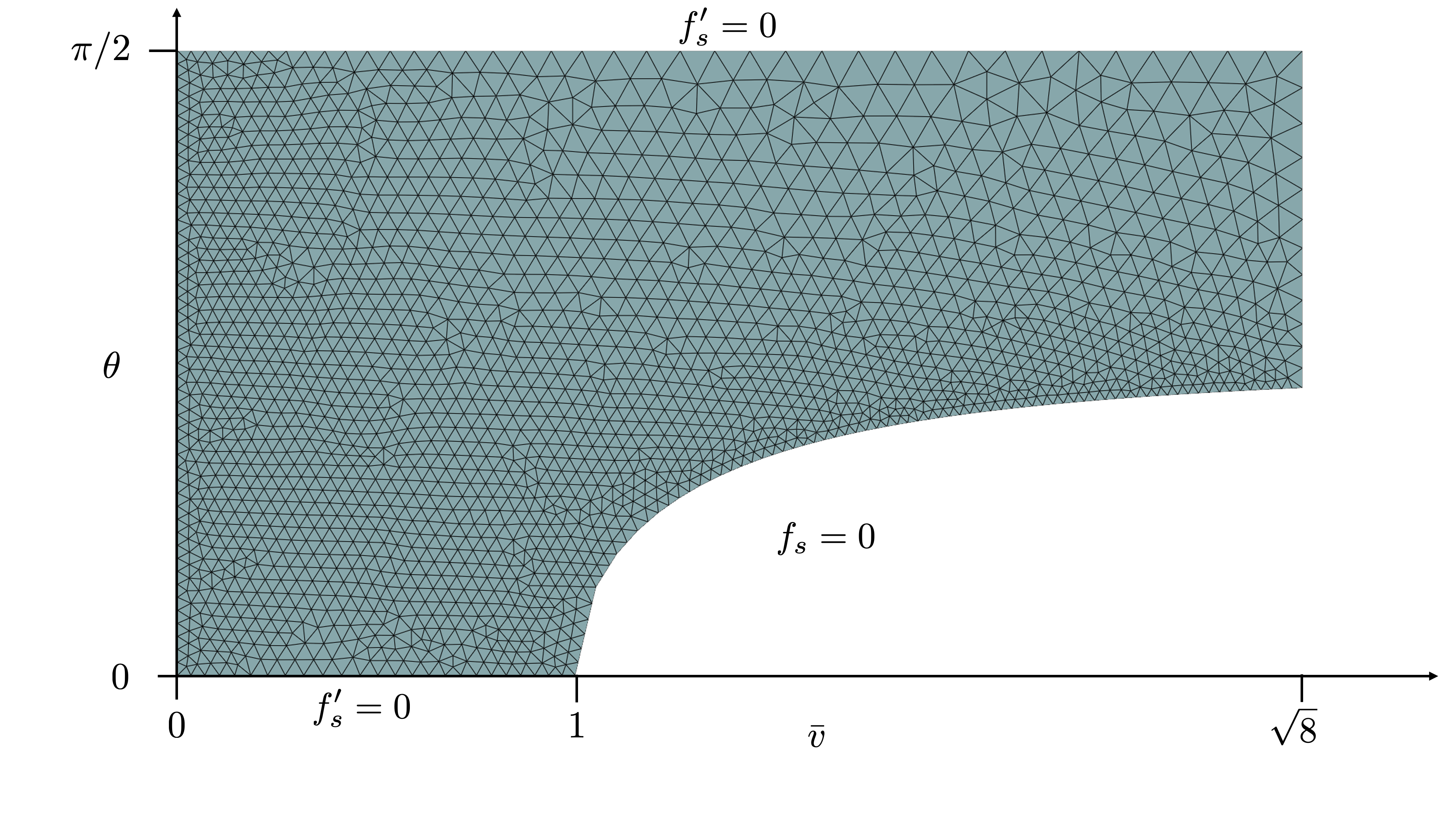}
    \caption{An example mesh from the finite element model. Here, $z_s e \phi / T_s = 1$ and $R = 2$ to exaggerate the loss cone.}
    \label{fig: FEM_mesh}
\end{figure}

\begin{figure}
    \centering
    \subfigure[Confinement time variations with ambipolar potential for $R = 10$ for electrostatically confined electrons $(Z_{s,N}=1)$.]{
        \includegraphics[height=1.8in]{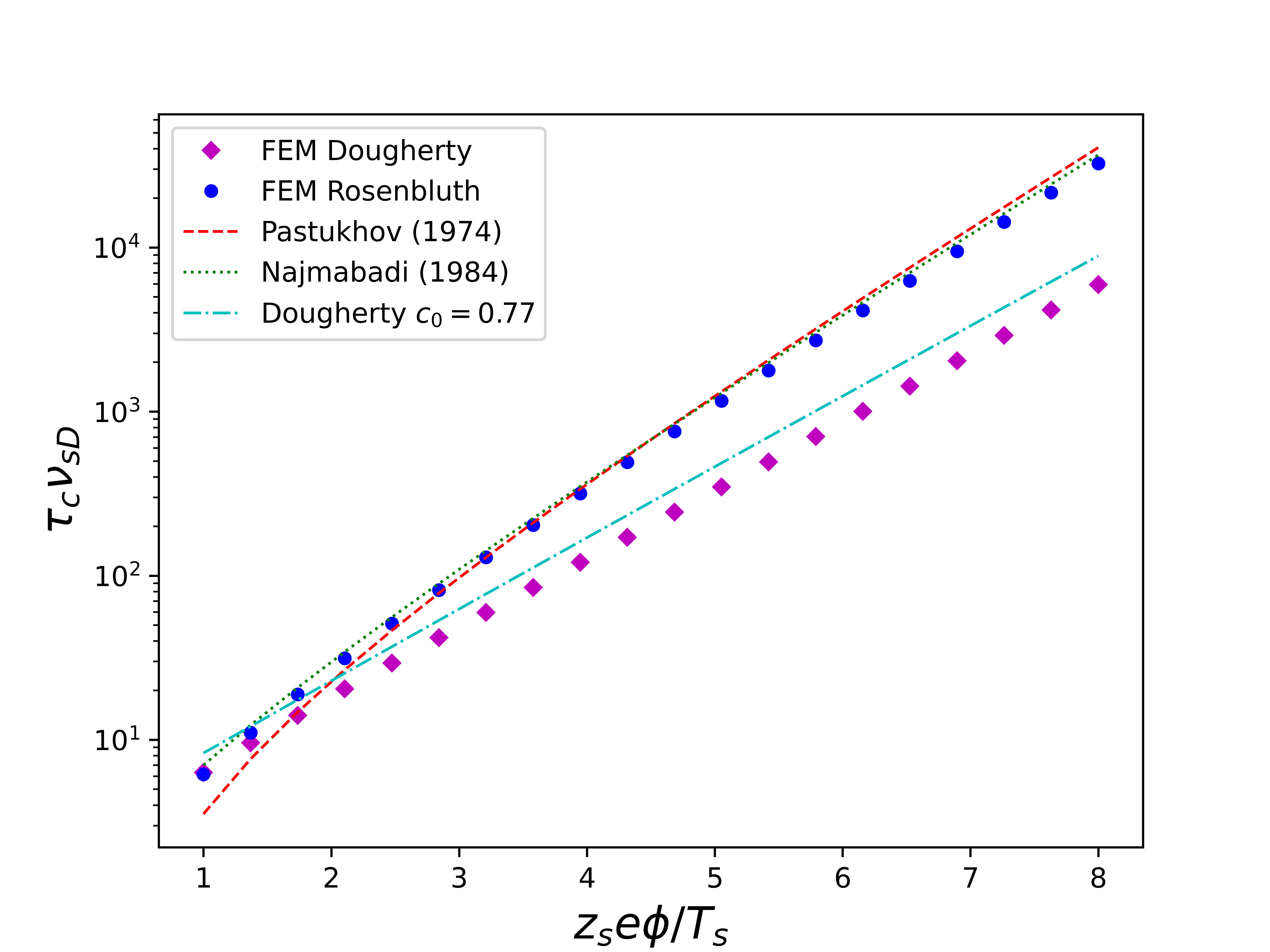}
        \label{fig: phi loss rate no correction}
    }
    \qquad
    \subfigure[Confinement time variations with mirror ratio for $z_s e \phi / T_s = 3$  for electrostatically confined electrons $(Z_{s,N}=1)$.]{
        \includegraphics[height=1.8in]{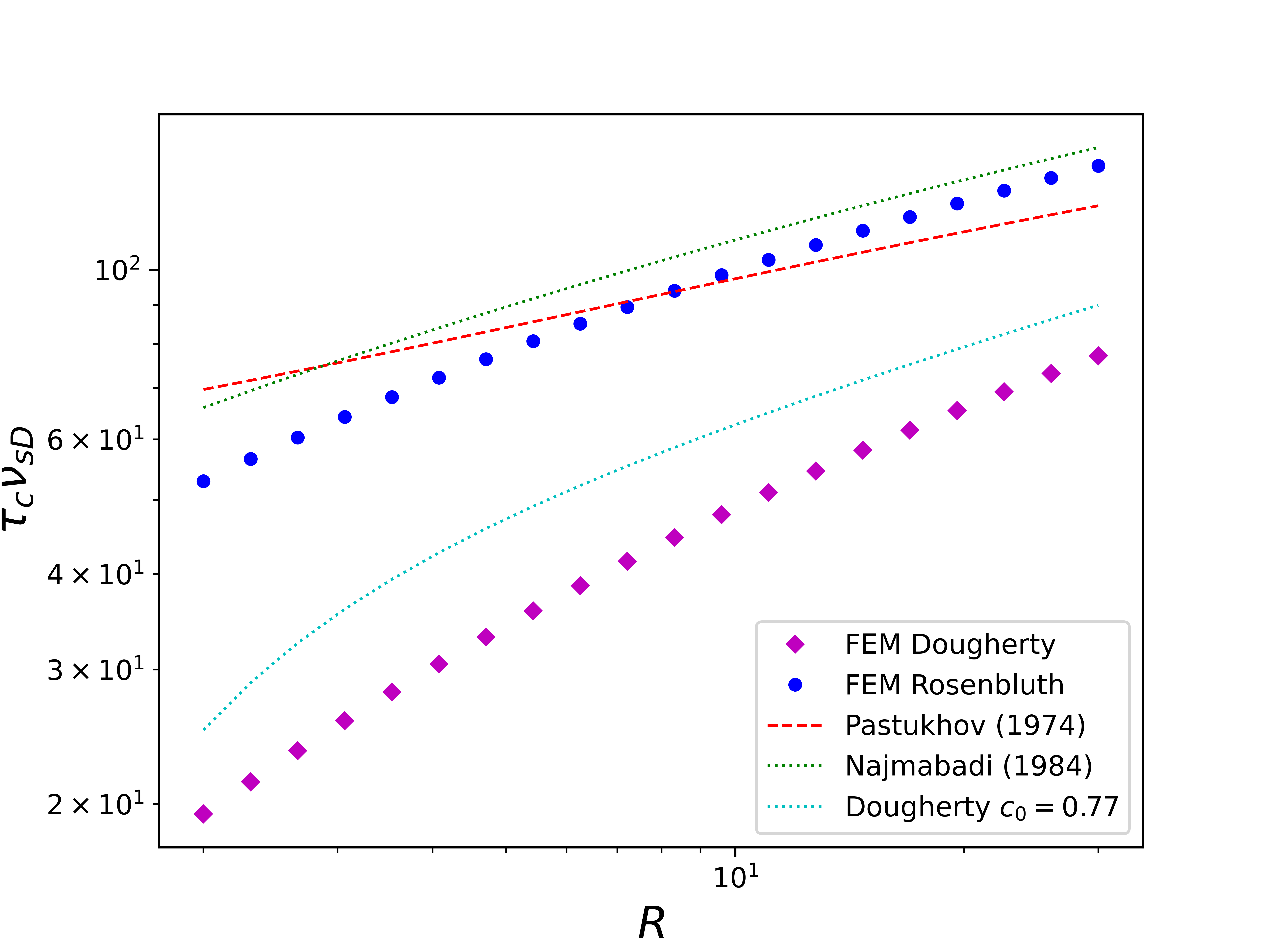}
        \label{fig: mirror ratio loss rate no correction}
    }
    \quad
    \subfigure[Variation of average energy of lost electrons $(Z_{s,N}=1)$ with ambipolar potential for $R=10$.]{
    \includegraphics[height=1.8in]{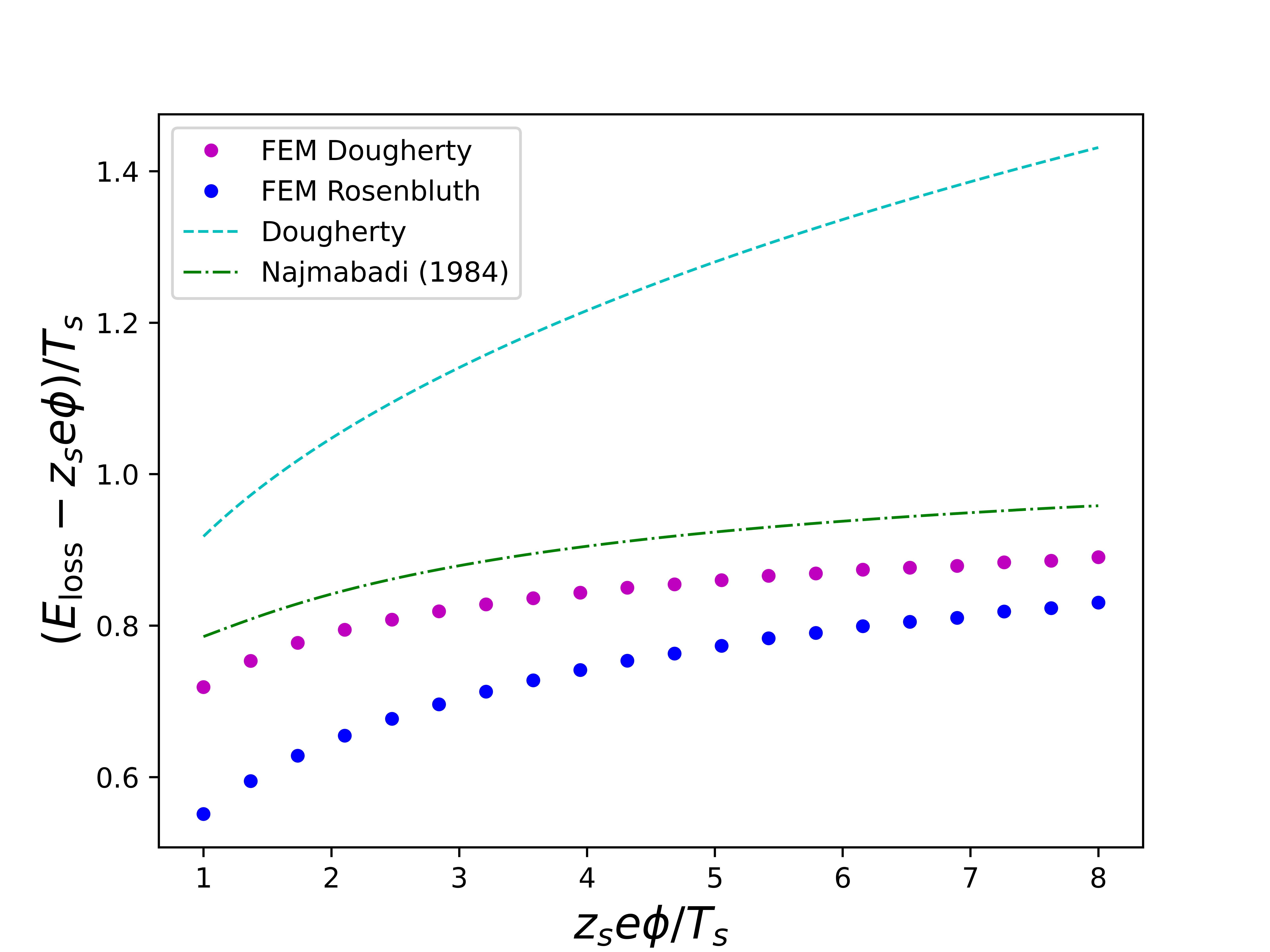}
    \label{fig: energy loss no corr}
    }
    \caption{\centering Normalized particle confinement time $\tau_c  \nu_{sLBO}$ from equation \eqref{eqn:elegant loss rates rosen} and its dependence on ambipolar potential and mirror ratio of the LBO versus \citet{pastukhov1974collisional}, with the correction noticed by \citet{cohen1978collisional}, and \citet{najmabadi1984collisional}. The y-axis is confinement time, normalized to the collision frequency with $Z_s = 1$ for electrostatically confined electrons.}
    \label{fig:comparison no correction}
\end{figure}

We compare our approximate expressions for the loss rates in magnetic mirrors to results obtained utilizing a code based on the work of \citet{ochs2023confinement}. The code uses the FEniCS DolfinX Python package to employ the finite element method (FEM) with third-order continuous Galerkin discretization to solve a general Fokker-Planck model collision operator. The FEM code is meshed over the upper right quadrant of $\bar{v},\theta$ space and has a low-energy source of the form $\bar{v}^2 \exp(-\bar{v}^2 / \bar{v}^2_{s0}) \theta^2 (\pi / 2 - \theta)^2$\footnote{There is a typo in Ochs (2023), where they miss a negative sign in the exponent.}. Here, $\bar{v}$ is normalized velocity, and $\theta$ is pitch angle. The normalized thermal velocity for the source is evaluated at $\bar{v}_{s0} = 0.2$ to concentrate it at low energy. This source form is chosen to go to zero on the boundaries smoothly (the precise form of the source term has little effect on the results in the asymptotic limit $z_s e \phi / T_e \gg 1$). Zero-flux boundary conditions are used at $\theta=0$, $\theta=\pi/2$, $v=0$, and $v=v_{max}$, the maximal velocity extent in the problem. A resolution of $\Delta \theta = \Delta \bar{v} = 0.1$ is chosen, with double resolution near the source and along the loss boundary. The boundary condition on the loss hyperboloid is Dirichlet ($f_s = 0$). The domain is extended $\sqrt{7 + z_s e \phi / T_s}$ past the loss hyperboloid vertex. An example mesh is shown in figure \ref{fig: FEM_mesh} to demonstrate. An astute reader will notice that the slope of the loss hyperboloid boundary is not vertical at $\bar{v} = 1$. At the tip of the loss hyperboloid, the code struggles to mesh the boundary because it is purely vertical. A finite step in $\bar{v}$ is taken, and the loss boundary is linearly interpolated from the tip of the loss hyperboloid to the first point. 
This plot was done with a coarser resolution to see the grid.  The final results were done at higher resolution confirmed to be adequate with convergence studies.
The problem is solved directly using PETSc's linear algebra solvers 
to find an LU decomposition.

To compare our numerical results to prior computational work, we need to make a key distinction between the collision operator utilized by our code in reproducing the results from \citet{najmabadi1984collisional} and the operators used in previous work. The results labeled ``FEM Rosenbluth'' utilize a slightly modified version of the collision operator in equation \eqref{eqn: Collision operator najmabadi}, whereas the code utilized in \citet{cohen1978collisional} and later utilized in \citet{najmabadi1984collisional} and \citet{fyfe1981finite} considers a multi-species Fokker-Planck equation derived from non-linear isotropic Rosenbluth potentials. Naive implementation of equation \eqref{eqn: Collision operator najmabadi} would violate the assumption that $\bar{v} \gg 1$ because our domain also includes low velocities. The work by \citet{cohen1978collisional} avoided this issue by implementing the full non-linear Rosenbluth potentials. To address this limitation, we approximate the parallel drag-diffusion frequency in equation \eqref{eqn: Collision operator najmabadi} by utilizing a Padé approximation of the form $1/\bar{v}^3 \rightarrow 1 /(1 + \bar{v}^3)$, which asymptotically matches the high- and low-velocity Rosenbluth limits. For the pitch angle scattering component, we have a full calculation of the pitch angle scattering rate while considering collisions on a Maxwellian background to keep the problem linear. We consider two cases: a plasma with hydrogen ions and electrons, where either the ions or electrons are electrostatically confined. In both cases, the thermal velocity of electrons is much higher than the thermal velocity of the hydrogen. This results in the following collision operator.
\begin{equation}
     \mathcal{L}_{\text{Code}}(F_s) = \frac{1}{\bar{v}^2} \frac{\partial}{\partial \bar{v}} \frac{\bar{v}^2}{1 + \bar{v}^3} \left( \bar{v} F_s + \frac{1}{2} \frac{\partial F_s}{\partial \bar{v}}\right) + \frac1{\bar{v}^3} P(v) \frac{\partial}{\partial \mu} \left( 1 - \mu^2 \right) \frac{\partial F_s}{\partial \mu}. \label{eqn: Collision operator code}
\end{equation}
For electrostatically confined electrons, $P(v) = (1 + \mathcal{R}(v))/2 $ ($Z_{s,n} = 1$). For electrostatically confined hydrogen ions, $P(v) = \mathcal{R}(v)/2$ ($Z_{s,n}=1/2$). In these expressions, $\mathcal{R}(v) = 1/(\sqrt{\pi}v) \exp(-v^2) + (1 - 1/2v^2)\rm{Erf}(v)$ and $\rm{Erf}$ is the error function. In contrast, the LBO model is implemented in the code without approximations as equation \eqref{eqn: Dougherty collision operator}. 

To demonstrate convergence, we examine increasing the resolution and the amount of the problem meshed beyond the tip of the loss hyperboloid for $z_s e \phi / T_s = 8$ and $R = 10$. When doubling all resolutions, the resultant loss rate changes by $-0.492\%$ for the Dougherty operator and $-1.099\%$ for the operator in equation \eqref{eqn: Collision operator code}. When not changing $\Delta \theta = \Delta \bar{v}$ but quadrupling the resolution near the source and along the loss hyperboloid, the loss rate changes by $-0.493\%$ for the Dougherty operator and $-1.001\%$ for the operator in equation \eqref{eqn: Collision operator code}. When increasing the extents of the problem from $\sqrt{7 + z_s e \phi / T_s}$ to $\sqrt{15 + z_s e \phi / T_s}$, the loss rate changes by $-0.06906\%$ for the Dougherty operator and $-0.06732\%$ for the operator in equation \eqref{eqn: Collision operator code}. Thus, the code converges within $1\%$ variation for all collision operators, and this resolution is suitable for this study.

In figure \ref{fig:comparison no correction}, the numerical code is validated by comparing the analytic results presented in \citet{pastukhov1974collisional} and \citet{najmabadi1984collisional} with corrections noticed by \citet{cohen1978collisional}, \citet{najmabadi1984collisional}, equation \eqref{eqn:elegant loss rates rosen}, and equation \eqref{eqn: average loss energy}, which in turn were validated using other numerical methods. Results in figure \ref{fig:comparison no correction} agree well with those in many prior works \citep{najmabadi1984collisional,cohen1978collisional,fyfe1981finite}.
The confinement time $\tau_c$ is normalized to collision frequency for generality. The diamonds are the finite element method numerical results, and the dashed lines are the analytic approximations. Although figure \ref{fig:comparison no correction} shows great agreement with prior work, it has a fair agreement with equation \eqref{eqn:elegant loss rates rosen}. To improve agreement between the code and equation \eqref{eqn:elegant loss rates rosen}, rather than calculating the flux correction coefficient by hand to be $0.77$ in \eqref{eqn: q0 bc matching with 0.77}, we can calculate this number numerically. First, let's call this correction coefficient $c_0$. It can be chosen to minimize error with the finite element code. Equations \eqref{eqn: q0 bc matching}, \eqref{eqn:elegant loss rates rosen}, and \eqref{eqn: energy loss rate} are modified to be the following.

\begin{align}
    \frac{1}{\tau_{c}} = \frac{1}{n_s} \frac{dn_s}{dt} &= -\nu_{sLBO} 2 Z_s \frac{ \mathrm{Erfc}(a)}{\ln \left( \frac{w+1}{w-1}\right) - c_0},\label{eqn: loss rate corrected}
\end{align}
\begin{align}
    \frac{1}{E_s}\frac{dE_s}{dt}= \frac{1}{n_s T_s} \frac{d (n_s T_s)}{dt} &= -\nu_{sLBO} \frac{4}{3 \sqrt{\pi}} Z_s  \frac{ e^{-a^2}a + \frac{\sqrt{\pi}}{2} \mathrm{Erfc}(a) }{\ln \left( \frac{w+1}{w-1}\right) - c_0}.\label{eqn: energy loss rate corrected}
\end{align}

\begin{figure}
    \centering
    \subfigure[Confinement time errors with ambipolar potential for $R = 10$.]{
        \includegraphics[height=1.8in]{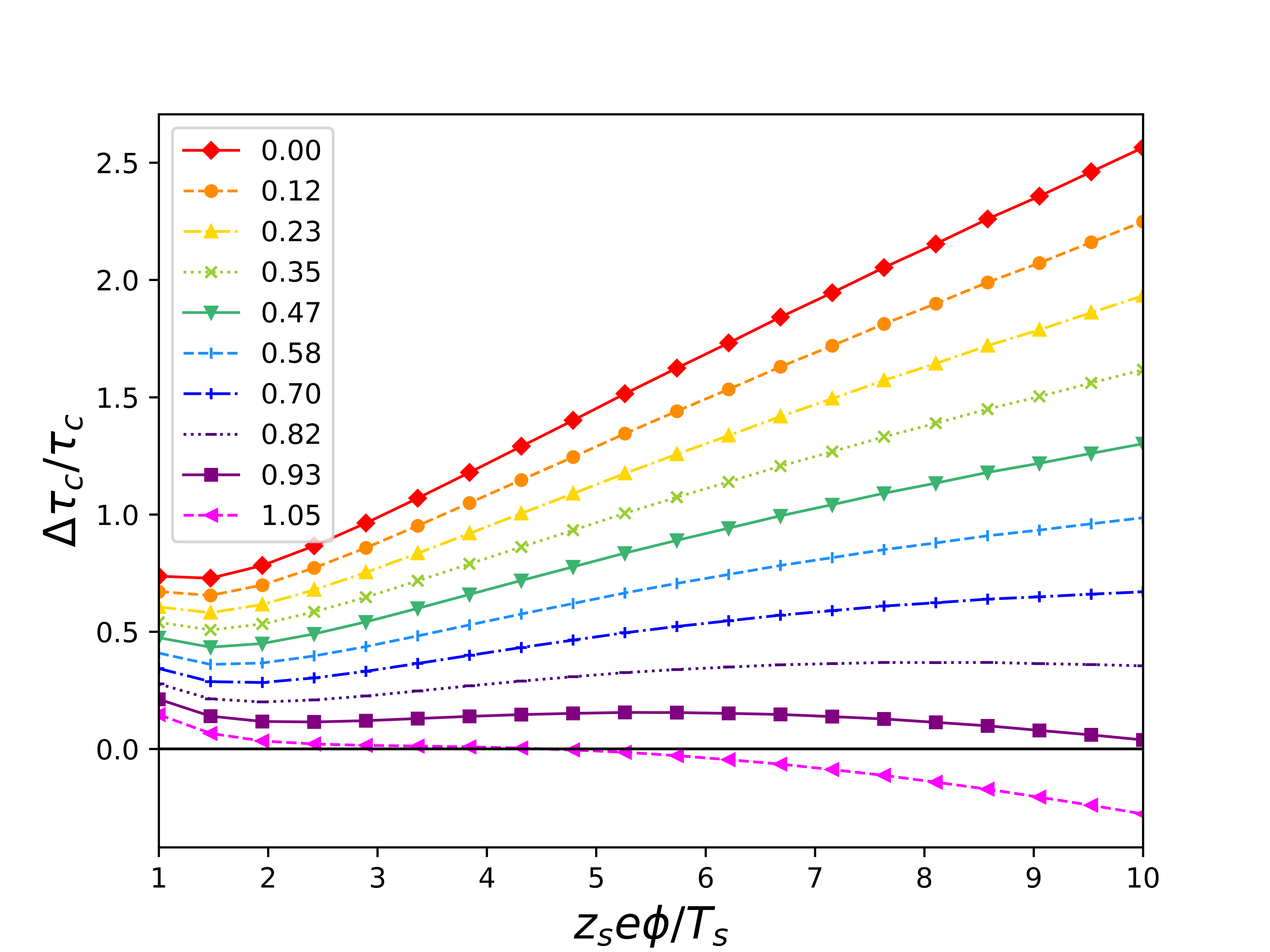}
        \label{fig: error phi loss rate}
    }
    \qquad
    \subfigure[Confinement time errors with mirror ratio for $z_s e \phi / T_s = 3$.]{
        \includegraphics[height=1.8in]{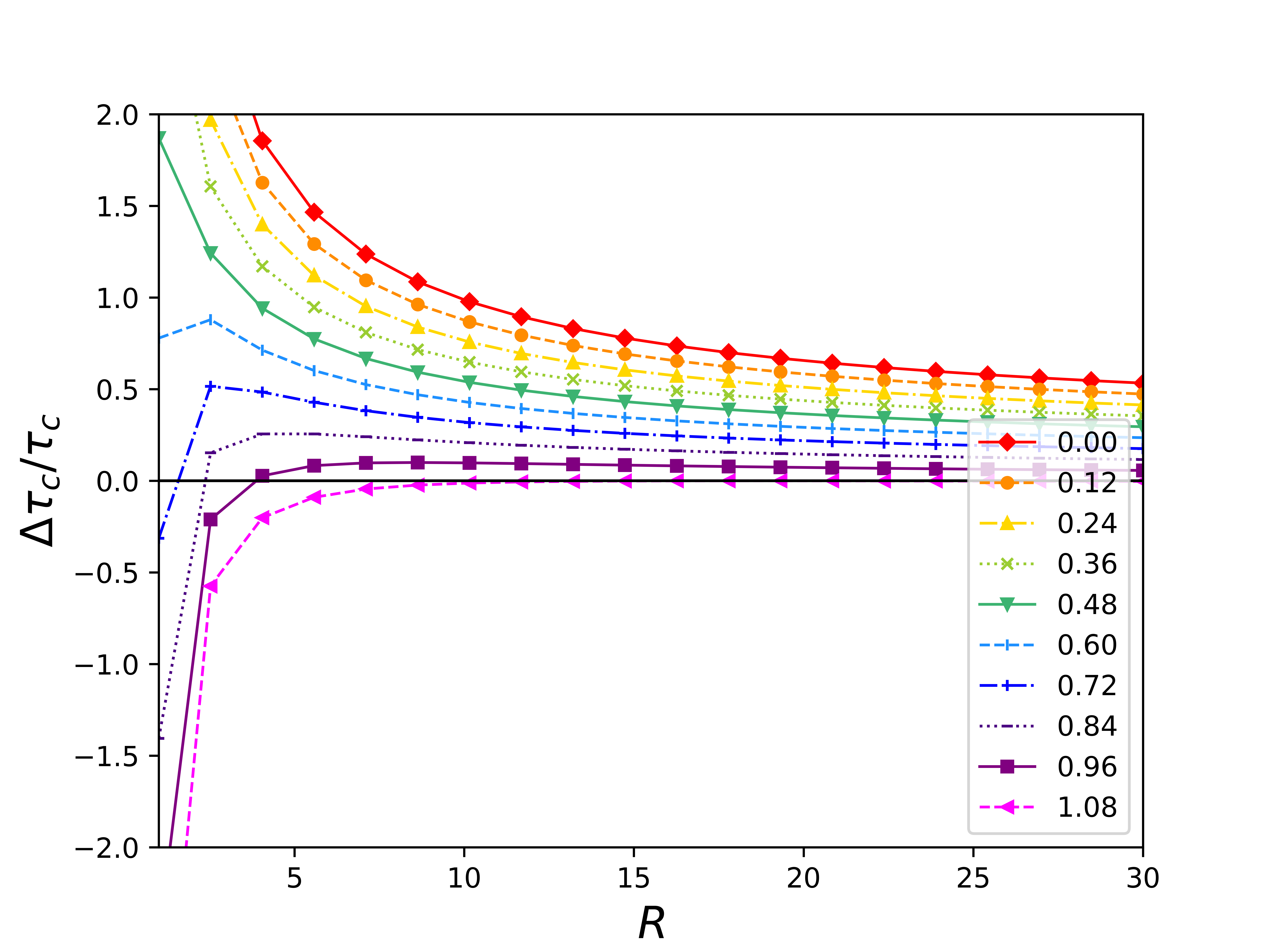}
        \label{fig: error mirror ratio loss rate}
    }
    \caption{\centering Fractional error in the confinement estimates between the numeric code and analytic approximation for electrostatically confined electrons. The legend goes from the top curve to the bottom curve in even steps in the value of $c_0$}
    \label{fig: error comparison}
\end{figure}

In figure \ref{fig: error comparison}, we show various curves that describe the error between equation \eqref{eqn: loss rate corrected} and the finite element code for various values of $c_0$. This figure shows us that the value of $c_0$ that minimizes error with the finite element code depends on the ambipolar potential and mirror ratio under investigation, although it seems to asymptote to a single value for a sufficiently large mirror ratio and ambipolar potential. In table \ref{tab: correction factors}, we list the appropriate correction factor to use for various values of ambipolar potential and mirror ratio that minimize error with the loss rate from the finite element code. 

\begin{table}
\centering
\begin{tabular}{@{}l *{10}{c}@{}}
& \multicolumn{10}{c}{$R$} \\ \cmidrule(lr){2-11}
$z_s e \phi / T_s$ &  5 & 10 & 15 & 20 & 25 & 30 & 35 & 40 & 45 & 50 \\
\midrule
1 & 1.305 & 1.310 & 1.302 & 1.295 & 1.288 & 1.283 & 1.278 & 1.274 & 1.271 & 1.268 \\
2 & 1.066 & 1.094 & 1.093 & 1.087 & 1.080 & 1.075 & 1.069 & 1.065 & 1.060 & 1.057 \\
3 & 1.004 & 1.066 & 1.078 & 1.080 & 1.077 & 1.074 & 1.070 & 1.066 & 1.062 & 1.059 \\
4 & 0.967 & 1.057 & 1.084 & 1.093 & 1.095 & 1.095 & 1.093 & 1.091 & 1.088 & 1.085 \\
5 & 0.933 & 1.044 & 1.083 & 1.099 & 1.107 & 1.109 & 1.110 & 1.109 & 1.108 & 1.106 \\
6 & 0.900 & 1.027 & 1.077 & 1.100 & 1.112 & 1.118 & 1.121 & 1.122 & 1.121 & 1.120 \\
7 & 0.869 & 1.007 & 1.065 & 1.093 & 1.109 & 1.117 & 1.122 & 1.125 & 1.126 & 1.126 \\
8 & 0.840 & 0.987 & 1.052 & 1.085 & 1.104 & 1.115 & 1.122 & 1.126 & 1.128 & 1.129 \\
9 & 0.813 & 0.968 & 1.038 & 1.077 & 1.099 & 1.113 & 1.122 & 1.128 & 1.132 & 1.134 \\
10 & 0.963 & 0.948 & 1.023 & 1.065 & 1.090 & 1.106 & 1.117 & 1.124 & 1.129 & 1.132 \\
\bottomrule
\end{tabular}
\caption{A table of optimal correction factors $c_0$ for various values of $z_s e\phi/T_s$ and $R$ to minimize error with the finite element code with $Z_s = 1$.}
\label{tab: correction factors}
\end{table}

\begin{figure}
    \centering
    \subfigure[Confinement time variations with ambipolar potential for $R = 10$, $c_0 = 1.04$, $Z_{s,N}=1$.]{
        \includegraphics[height=1.8in]{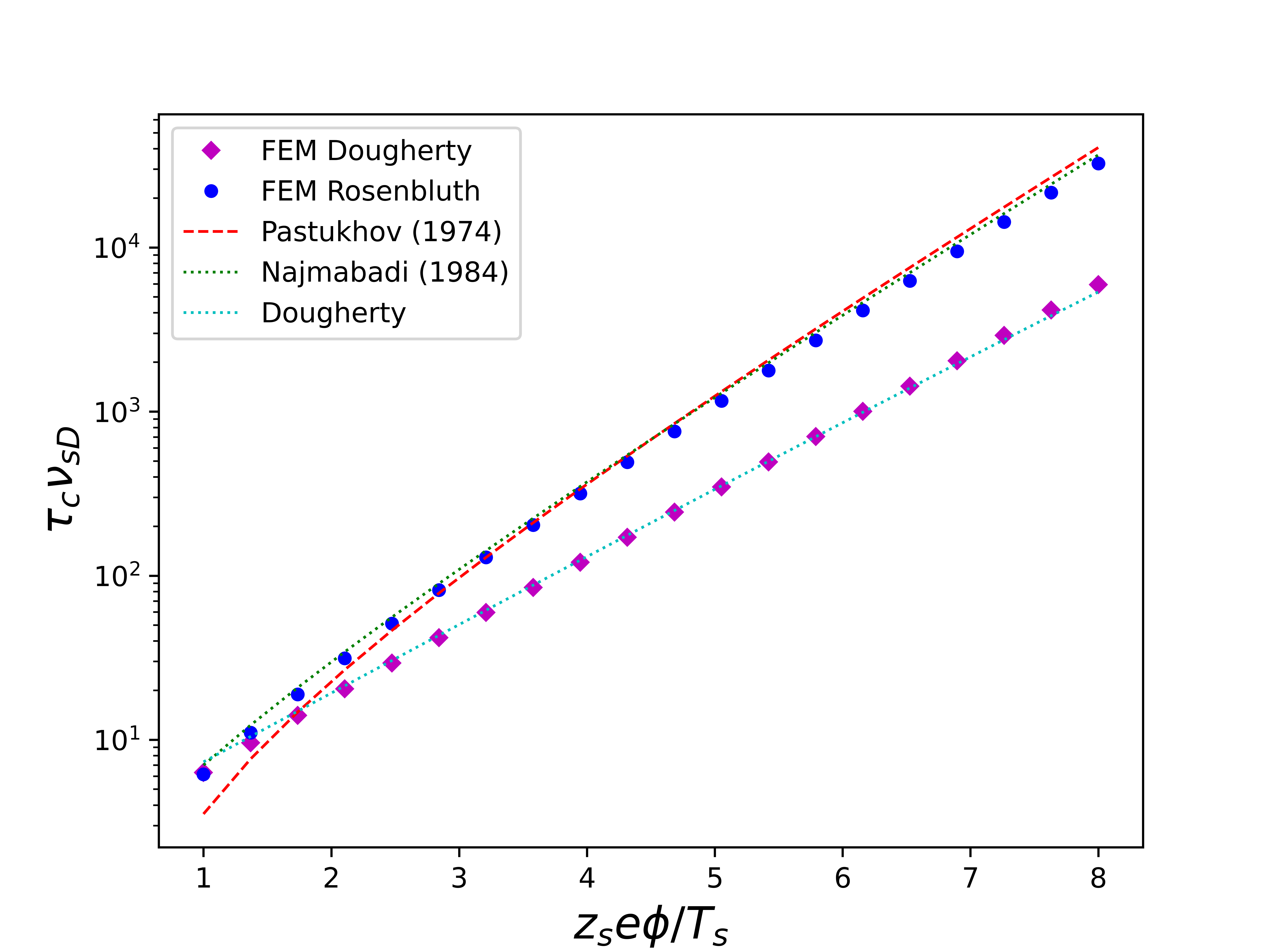}
        \label{fig: phi loss rate}
    }
    \qquad
    \subfigure[Confinement time variations with mirror ratio for $z_s e \phi / T_s = 3$ and $c_0 = 1.07$, $Z_{s,N}=1$.]{
        \includegraphics[height=1.8in]{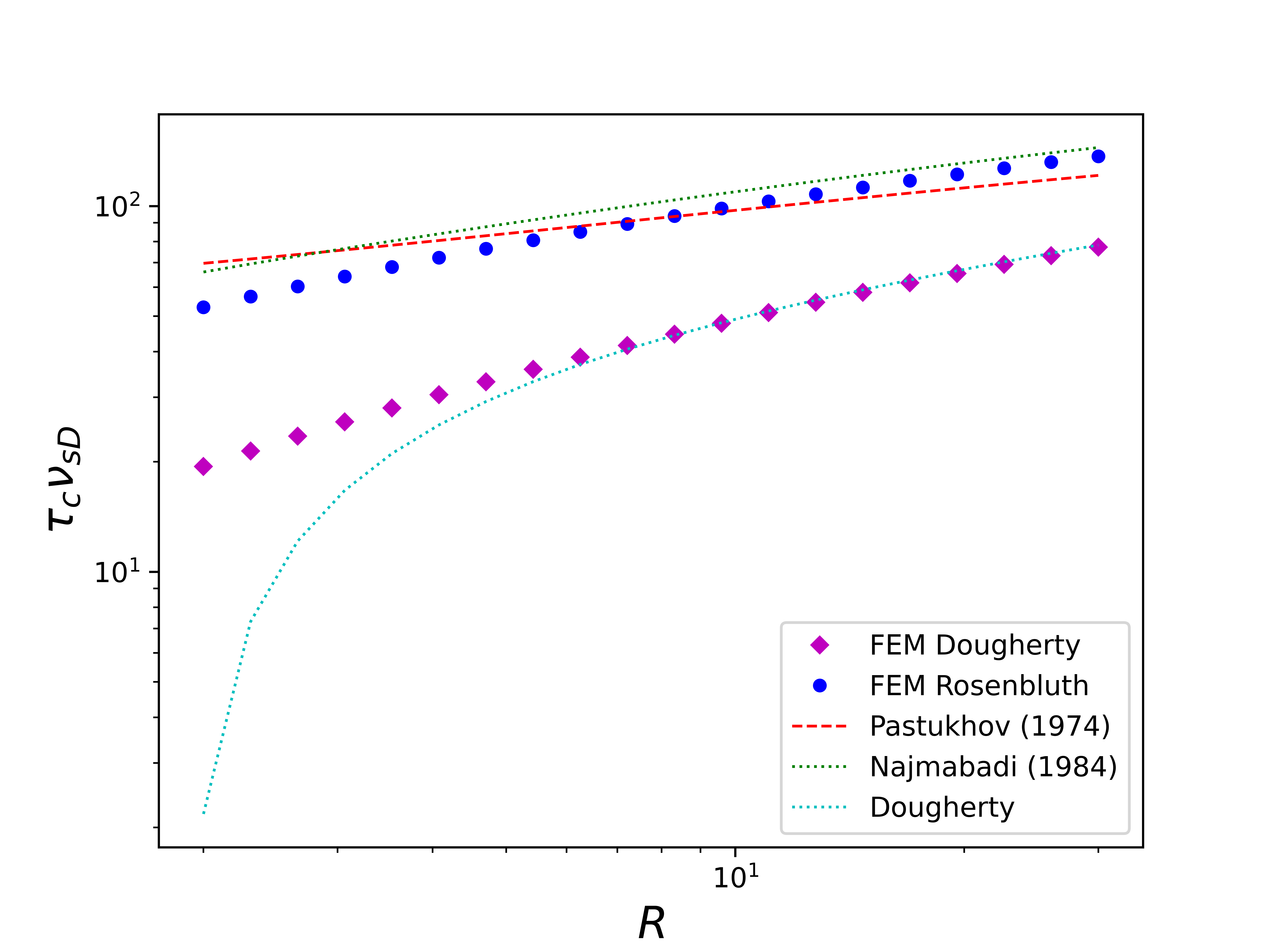}
        \label{fig: mirror ratio loss rate}
    }
    \quad
    \subfigure[Variation of average energy of lost particles with ambipolar potential for $R=10$, $c_0 = 1.04$, $Z_{s,N}=1$.]{
    \includegraphics[height=1.8in]{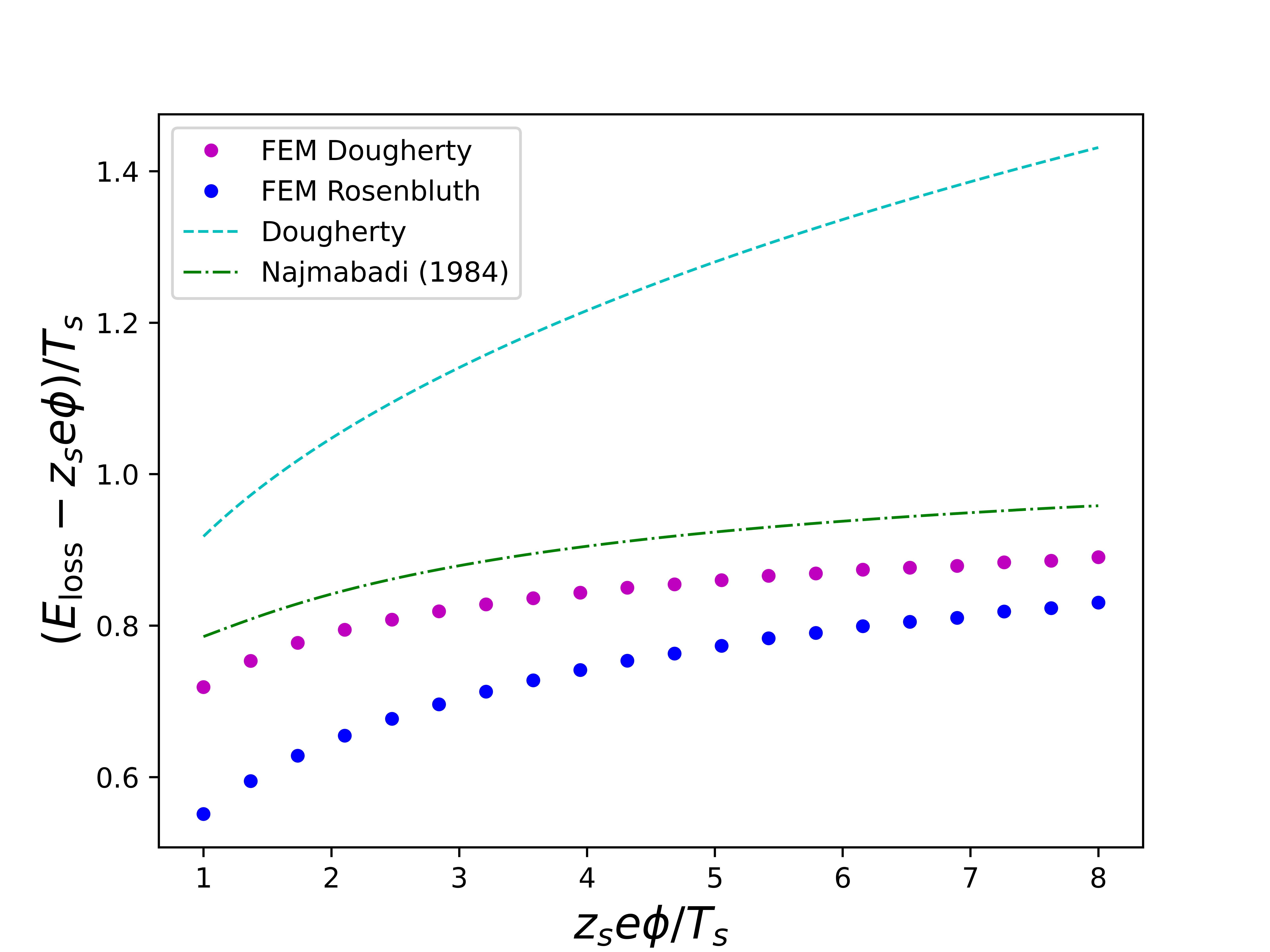}
    \label{fig: energy loss}
    }
    \caption{\centering Comparison particle confinement time and average loss energy, and its dependence on ambipolar potential and mirror ratio of the LBO versus \citet{pastukhov1974collisional} and \citet{najmabadi1984collisional}. The y-axis is confinement time or average loss energy subtracted by $z_s e\phi / T_s$, normalized to the collision frequency with $Z_s = 1$ for electrostatically confined electrons.}
    \label{fig:comparison}
\end{figure}

Figure \ref{fig:comparison no correction} is reconstructed in figure \ref{fig:comparison} using the correction coefficient $c_0$ in equations \eqref{eqn: loss rate corrected} and \eqref{eqn: energy loss rate corrected}. In figure \ref{fig: energy loss}, a choice to not plot the results from \citet{pastukhov1974collisional} is made due to their close agreements with the results from \citet{najmabadi1984collisional}. 

In comparing the dependence of the loss rate with ambipolar potential, figure \ref{fig: phi loss rate} shows a key difference between the Fokker-Planck form Coulomb operator and the LBO. Using a constant mirror ratio of $R=10$, the LBO alters the confinement time compared to the Fokker-Planck collision operator utilized in other studies \citep{khudik1997longitudinal,najmabadi1984collisional,pastukhov1974collisional,catto1981collisional,catto1985particle}. Particularly, the confinement time of \citet{najmabadi1984collisional} scales as $a^2 e^{a^2}$, but equation \eqref{eqn: loss rate corrected} scales as $a e^{a^2}$. This scaling difference makes sense because the LBO overestimates the collision frequency and ignores the collisionality drop-off with velocity. 

Figure \ref{fig: phi loss rate} and \ref{fig: mirror ratio loss rate} compare the variation with potential and mirror ratio of the loss ratios calculated using different collision operators. We see that the LBO model underestimates the confinement time at all potentials and mirror ratios. Even so, the percentage variation of loss rate upon modification of the mirror ratio has similar trends in the finite element code when comparing the LBO model to equation \eqref{eqn: Collision operator code}. Furthermore, validating the code, each numerical line matches their respective analytic curves well. Figure \ref{fig: mirror ratio loss rate} leads us to conclude that there is no significant difference in the dependence of loss rate on mirror ratio when considering LBO collisions.

Figure \ref{fig: energy loss} shows that our equation \eqref{eqn: average loss energy} and the numerical approach exhibit similar trends, but are very different in magnitude. It is important to note that figure \ref{fig: energy loss} subtracts the linear component $z_s e \phi / T_s$ to highlight the differences between the different results. Comparing results to prior work, our numerical method produces roughly a 20\% difference with respect to the analytic results from \citet{najmabadi1984collisional}, investigated in appendix \ref{sec: Najmabadi correction}. Although errors of 20\% are large, \citet{najmabadi1984collisional} noticed similar errors, so these results are comparable to prior work. Interestingly, while the analytic results of \citet{najmabadi1984collisional} seem to have a constant 20\% error, the discrepancy between \eqref{eqn: average loss energy} and the numerical results grows with $z_s e \phi / T_s$. With the LBO at $c_0 = 0.77$, similar errors of 20\% are observed at low values of $z_s e \phi / T_s$; at large values, the error grows to nearly 50\% at $z_s e \phi / T_s = 8$. The LBO model has an overall higher average loss energy than the model used by \citet{najmabadi1984collisional}. This means the losses from LBO collisions are more spread around the loss hyperboloid. In contrast, a more accurate collision operator would have losses more concentrated around the tip of the loss hyperboloid. It is worth noting that the correction factor $c_0$ does not modify equation \eqref{eqn: average loss energy}.

%% file: discussion.tex
\section{Discussion}
\label{sec: discussion}

We suggest that codes using an LBO/Dougherty collision operator improve their results by scaling the collision frequency used to obtain the correct mirror confinement time and ambipolar potential. First, one must obtain an accurate estimate of the ambipolar potential of the system. The ambipolar potential may be determined using the analytic results presented in \cite{najmabadi1984collisional} or in a code such as our finite element solver or the one presented in \cite{egedal2022fusion}. Using this accurate estimate of ambipolar potential, one can calculate the estimated confinement time using the results from \citet{najmabadi1984collisional}, presented in appendix \ref{sec: Najmabadi correction}. Call this confinement time $\tau_N$. Now that we have calculated the ambipolar potential and confinement time, we may invert equation \eqref{eqn: loss rate corrected} to determine $\nu_{sLBO} Z_s$, with the appropriate correction coefficient using table \ref{tab: correction factors},
\begin{equation}
    \frac{1}{\nu_{sLBO}Z_s} = 2 \tau_{c,N} \left( \ln \left( \frac{w+1}{w-1}\right) - c_0\right)^{-1} \mathrm{Erfc}(a), \label{eqn: invert dougherty loss rate}
\end{equation}
Here, recall that $w = \sqrt{1 + 2 z_s e \phi / \left( T_s Z_s R \right)}$, and $a = \sqrt{ z_s e\phi / T_s + \ln{(w)}}$. The simplest approach to achieving the correct results is to scale the collision frequency, setting $Z_s = 1$, which will also have undesirable effects such as modifying multi-species thermal equilibration times \citep{francisquez2022improved}. Let us define a collision frequency gain constant $\gamma = \nu_{sLBO} / \nu_{N}$ where $\nu_N$ is the collision frequency used in the results of \citet{najmabadi1984collisional}. One must scale the collision frequency $\nu_{N}$ by a factor of $\gamma$ for the loss rate of the system using the LBO to be equal to the system using the operator from \citet{najmabadi1984collisional}. Alternatively, one could implement an LBO with a modified diffusion coefficient, including $Z_s$. Then, one may scale $Z_s$ by a factor of $\gamma$ instead of collision frequency, which may be preferable depending on the situation. $Z_s \ne 1$ corresponds to an anisotropic collision operator, which can be more accurate but, in some cases, can be numerically more challenging \citep{sharma2011fast}.

Putting this into practice, we make suggestions of the appropriate scaling factor $\gamma$ for WHAM and WHAM++. \citet{egedal2022fusion} predicts an ambipolar potential of $z_s e\phi/T_e \simeq 5$ for WHAM and WHAM++. For WHAM, $R = 13.3$ and $\gamma = 0.3030$, and for WHAM++, $R=10$ and $\gamma = 0.2721$ with $c_0=1.05$ and $Z_s=1$ for both machines. It makes sense that the LBO would need a diminished frequency to match Najmabadi's loss rate because of the $1/v^3$ drop-off in collision frequency at large velocity.

The scrape-off layer of toroidal confinement devices, like tokamaks and stellarators, can exhibit a Pastukhov potential due to the difference in magnetic field strength between the inboard and outboard sides \citep{Majeski2017}. The magnetic field magnitude ($|B|$) varies inversely with the distance ($R$) from the central axis, determining the mirror ratio ($R_\mathrm{out}/R_\mathrm{in}$) of these devices, typically around 2. Additionally, observed potentials are approximately $z_s e \phi / T \sim 2$. This implies these devices have low mirror ratios and a limited potential. In this scenario, figure \ref{fig: phi loss rate} aligns reasonably with previous findings by \citet{najmabadi1984collisional}. However, \citet{pastukhov1974collisional} and \citet{najmabadi1984collisional} consider a high mirror ratio and potential limit, making their results inappropriate for toroidal confinement devices. Caution is necessary when applying these equations in regimes approaching the bounds of these limits. The variational method for studying collisional losses not considered here offers an approach suitable for arbitrary mirror ratios and ambipolar potential, making results from \citet{catto1985particle} and \citet{khudik1997longitudinal} more appropriate. 


We note a subtlety in the problem setup as defined by Pastukhov, Najmabadi, and others.
It is usually stated that there is a low-energy source of particles to balance the sink at high energy.  
Standard electron-electron collision operators are energy conserving, and the electron-ion collision operator has been approximated by pitch angle scattering, which also conserves energy, so one may wonder where the energy injection comes from.  
It should be noted that it is not sufficient to add a source to the kinetic equation of the form $S(v) = S_0 \delta(v-v_I) / (4 \pi v^2)$ (where $S_0$ is the source rate in units of particles per second per unit volume), because no choice of the injection velocity $v_I$ can provide enough power for steady state unless $v _I$ is quite large, violating the assumption that the source is at low energy.  
This is because the source energy per injected particle, $(1/2) m v_I^2$, must be the same in steady state as the average energy per particle lost, which from \eqref{eqn: average loss energy asymptotic} is $E_{\rm loss} \sim z_s e \phi  \gg T_e$.  

The resolution to this issue is as follows: in a real mirror machine, the heating of electrons comes from collisions with beam and bulk ions (typically much hotter than the electrons) or RF heating.  Collisions and quasilinear RF heating can be roughly modeled here by slightly increasing the velocity diffusion coefficient in the collision operator, i.e. slightly increasing the factor of $v_{{\rm th},s}^2$ that appears in the Dougherty-Lenard-Bernstein collision operator (\eqref{eqn:LBO}) above the actual thermal velocity squared $\langle v^2 \rangle / 3$ one would find from distribution function. 
Because the mirror confinement time $\tau_c$ is much longer than the collision time in the asymptotic regime we are studying, roughly by a factor of $\sim \exp(a^2) \sim \exp(z_s e \phi / T_e) \gg 1$ (neglecting factors of $a$ or $a^2$ depending on the choice of collision operator), the velocity diffusion coefficient only needs to be increased slightly (the relative change scales as $\exp(- z_s e \phi / T_e) \ll 1$). 
In numerical codes, this is sometimes implemented via a hidden assumption by not literally using an energy-conserving electron-electron collision operator and instead fixing the value of $v_{{\rm th},s}^2$ in the diffusion coefficient as an input parameter.  One would find that the actual $\langle v^2 \rangle / 3$ from the distribution function is slightly less than the input parameter.
Some calculations do the same thing by normalizing the velocity variable to a fixed parameter, which will turn out to be slightly higher than the actual thermal velocity.
Nevertheless again, this is a tiny correction in the asymptotic limit.

%% file: conclusion.tex
\section{Conclusion}\label{sec: conclusion}

In summary, this study delves into the critical role that collisions play in governing particle and energy transport within magnetic mirror confinement systems. We use an LBO model to proceed through the method of images calculation to investigate particle and energy confinement. Notably, we address the challenges posed when using a different collision operator compared to prior work. 

A pivotal observation emerges when examining the dependence of confinement time on the ambipolar potential, as depicted in figure \ref{fig: phi loss rate}. The LBO alters the particle confinement time compared to more accurate collision operators utilized in earlier research \citep{khudik1997longitudinal,najmabadi1984collisional,pastukhov1974collisional,catto1981collisional,catto1985particle}. Notably, our findings demonstrate that the particle confinement time scales like $a \exp(a^2)$ using the LBO, whereas a more accurate collision operator would yield $a^2 \exp(a^2)$, where $a^2$ is approximately the normalized ambipolar potential, $z_s e \phi / T_e$ (see the inline equations after equation \eqref{eqn: energy loss rate}). This scaling discrepancy is attributed to the LBO's disregard for the drop-off in collisionality with velocity. Figure \ref{fig: mirror ratio loss rate} highlights that despite significantly different loss rates at the same potential, the scaling behavior with the mirror ratio remains comparable with prior models. Finally, figure \ref{fig: energy loss} showcases that equation \eqref{eqn: energy loss rate corrected} reproduces comparable errors compared to prior work.

To address these findings, we propose a practical modification for codes utilizing an LBO or Dougherty operator to achieve the correct ambipolar potential. This involves estimating the particle loss rate, then calculating the ambipolar potential from a code or predictions from \citet{najmabadi1984collisional}. Equation \eqref{eqn: loss rate corrected} can then be employed to determine the appropriate scaling factor for the collision frequency or pitch angle scattering enhancement, ensuring congruence with the electron confinement time. This scaling factor, denoted as $\gamma$, depends on various parameters, including the correction factor $c_0$, which can be interpolated from table \ref{tab: correction factors}.

Numerous avenues for extending this research exist. Firstly, the incorporation of anisotropic diffusion coefficients has the potential to alleviate the approximations inherent in the LBO operator. This is facilitated by the presence of $Z_s$ in equations \eqref{eqn: loss rate corrected} and \eqref{eqn: energy loss rate corrected}. Additionally, we have expanded the method of images approach to accommodate alternative forms of the Fokker-Planck coefficients used in the collision operator. Future investigations could explore alternative approximate collision operators using this generalized approach or even take the approximations from \citet{najmabadi1984collisional} to higher order. In addition, the code could be improved by including an improved approximation to the energy diffusion term in equation \eqref{eqn: Collision operator code} using the Rosenbluth potentials. Furthermore, future research endeavors might delve into variational techniques, as demonstrated by \citet{catto1981collisional}, \citet{catto1985particle}, and \citet{khudik1997longitudinal}, to relax the constraints imposed by large mirror ratios within the method of images. On the computational side, it would be interesting to see how much the results of \citet{francisquez2023towards} change with rescaling the collision frequency suggested here. Extending the lessons learned here beyond magnetic mirrors, it is worth considering tokamak and stellarator confinement systems. As discussed, toroidal confinement systems exist in the space of low mirror ratios and low potentials, which we studied to be a regime where the loss rate due to LBO collisions has a reasonable agreement with more comprehensive collision operators, and little modification needs to be made. However, this analysis assumed large mirror ratios and large potentials, so more careful analysis must be done using variational techniques for a complete answer.

%% file: appendix.tex
\appendix
\section{Convenient choice for the free function $q(\bar{v})$}
\label{sec: Appendix 1}
We will prove equation \eqref{eqn: q form} by first examining the delta function $\delta(1-\mu^2)$ to transform it in terms of $\rho$. Recalling equation \eqref{eqn: rho transformation}, we reorganize this definition to have equality
\begin{equation}
    \frac{Z_s \mu^2 \rho^2}{4 \bar{v}^4} e^{-2 \bar{v}^2} =   1 - \mu^2.
\end{equation}
Using the delta function to assert that $\mu=\pm 1$, we can show that the delta function transforms to 
\begin{align}
    \delta\left(1-\mu^2\right) = \delta\left(\frac{Z_s \rho^2}{4 \bar{v}^4} e^{-2 \bar{v}^2}\right) = \delta\left(\rho\right) \frac{2 \bar{v}^4}{Z_s \rho} e^{2 \bar{v}^2},
\end{align}
where we have used $\delta(g(x)) =  \delta(x-x_0)/|g'(x_0)|$ with $x_0$ satisfying $g(x_0) = 0$.
Transforming $Q(\bar{v},\mu)$, we then get 
\begin{align}
    \frac{\pi^{3/2}}{2 z \ln(z)} \frac{\delta(1-\mu^2)}{4\pi} H(\bar{v}-a) q(\bar{v}) = \frac{\delta\left(\rho\right)}{2 \pi \rho} H(z - z_a) 4 \pi \left( \frac{ z \ln(z) \sqrt{\pi}}{8  Z_s }    q(\bar{v}) \right),
\end{align}
where we have grouped terms to get it in the form of equation \eqref{eqn: Poisson source form}. We find equation \eqref{eqn: q form} by setting $q(\bar{v})$ to cancel out this functional dependence and leave within it some arbitrary functionality $\bar{q}$. This follows the approach set forth by \citet{najmabadi1984collisional}, where they set this to a constant for simplicity.

\section{Investigation into the correction factor in \citet{najmabadi1984collisional}}
\label{sec: Najmabadi correction}
The notation used in this paper is very similar to the notation used in \citet{najmabadi1984collisional}. For clarity, let us define some of the key differences in notation, which will be used exclusively in this appendix section. For the collision operator used in \citet{najmabadi1984collisional}, they define the appropriate collision frequency and anisotropic diffusion coefficients considering multispecies collisions.
\begin{align}
    \nu_e &\equiv \frac{4 \pi}{m_e^2 v_{th,e}^3} \left(\frac{e^2}{4 \pi \epsilon_0}\right)^2 n_e \lambda_{ee}, && Z_{e,N} \equiv \frac{1}{2} \left(1+ \frac{\sum_j n_j z_j^2 \lambda_{ej}}{n_e \lambda_{ee}} \right),\\
    \nu_i &\equiv 4 \pi \left( \frac{e^2}{4 \pi \epsilon_0} \right)^2\sum_j \frac{n_j z_i^2 z_j^2 \lambda_{ij}}{m_i m_j v_{th,i}^2} \frac{T_j}{T_i},&& Z_{i,N} \equiv \frac{1}{2} \frac{\sum_j n_j z_j^2 \lambda_{ij}}{\sum_j n_j z_j^2 \lambda_{ij} (T_j / T_i) (m_i / m_j)}.
\end{align}
where $\sum_j$ is a summation over ions only, $v_{th,s} = \sqrt{2 T_s/m_s}$ $e$ is the electronic charge, $z_s$ is the atomic number of species $s$, $\lambda_{ab}$ is the Coulomb logarithm, and $\epsilon_0$ is the dielectric constant. 

We restate the main finding from \citet{najmabadi1984collisional}. They found that the confinement time scales as
\begin{equation}
    \frac{1}{\tau_{c, N}} = \nu_s \frac{4}{\sqrt{\pi}} \frac{\exp(-a^2)}{a^2} \frac{ \left( Z_{s,N} + 1/4 \right) a^2 \exp(a^2) E_1(a^2) - 1/4 }{\ln \left( \frac{w + 1}{w - 1}\right) - c_N}. \label{eqn: Najmabadi loss rate result}
\end{equation}
Here, $ w = \sqrt{1 + 2 / \left( R (Z_s-T_s/4 z_s e\phi)\right)}$, and $a^2 = z_s e \phi / T_s + \ln(w)$, $E_1(x) = \int_x^\infty \rm{dt}\exp(-t)/t $ is the exponential integral, $c_N$ is a correction coefficient, and the subscript $N$ stands for \citet{najmabadi1984collisional},\footnote{There is a typo in the second statement of $w$ in \citet{najmabadi1984collisional} near equation (42) which should have a square root, noticed by \citet{ochs2023confinement}.}. 

\begin{figure}
    \centering
    \subfigure[Confinement time errors with ambipolar potential for $R = 10$.]{
        \includegraphics[height=1.8in]{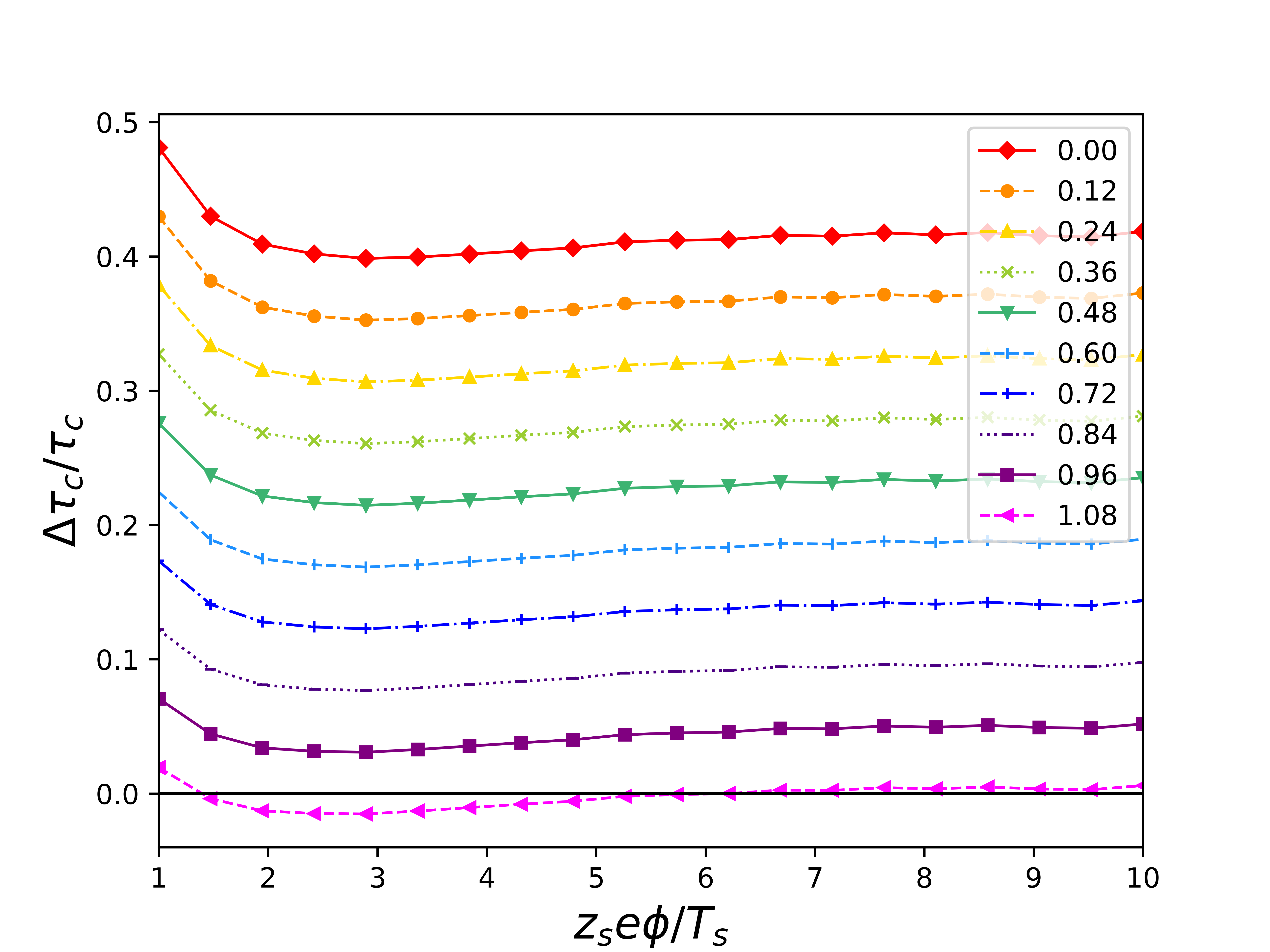}
        \label{fig: error phi loss rate}
    }
    \qquad
    \subfigure[Confinement time errors with mirror ratio for $z_s e \phi / T_s = 3$.]{
        \includegraphics[height=1.8in]{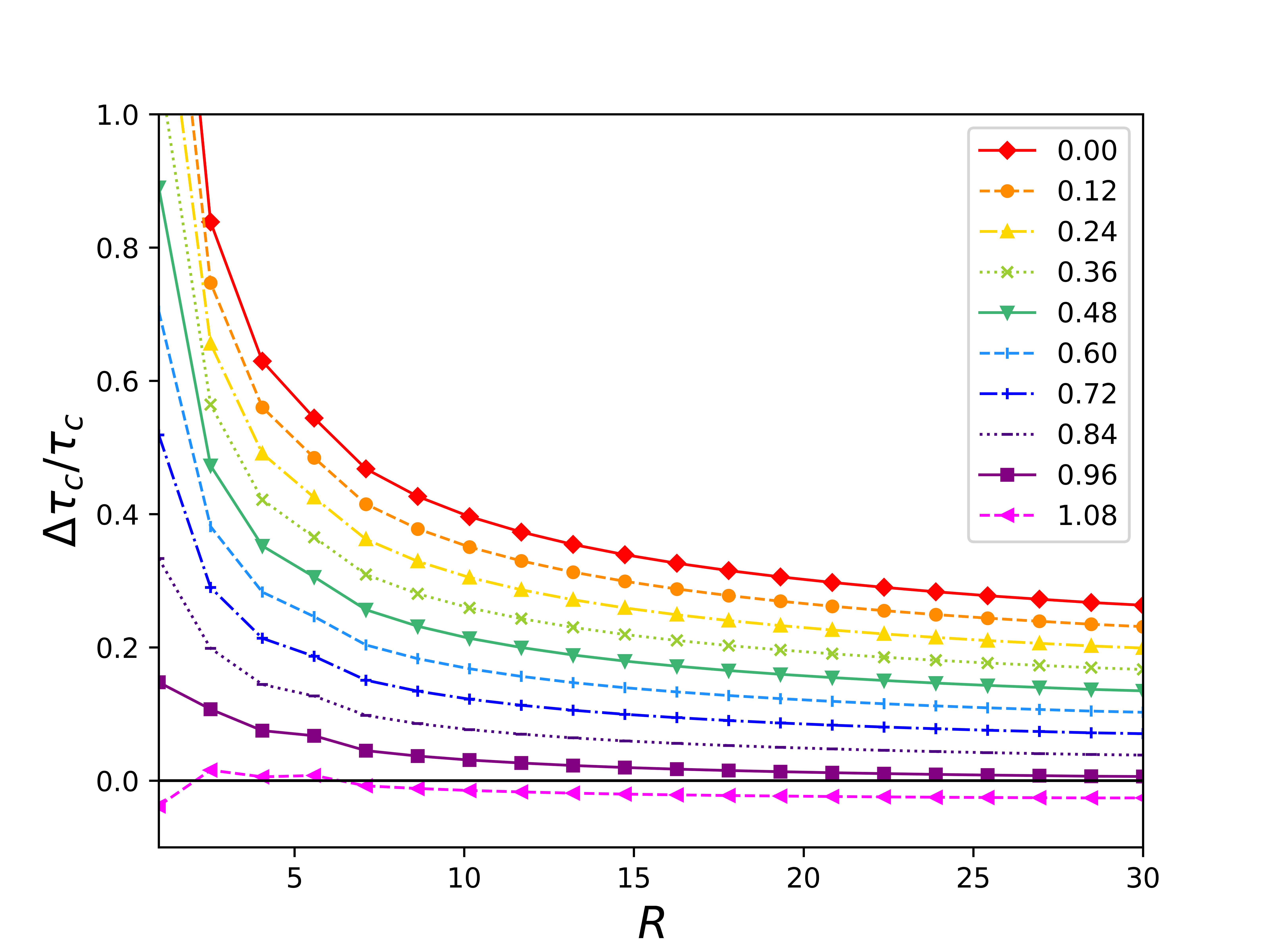}
        \label{fig: error mirror ratio loss rate}
    }
    \caption{\centering Fractional error in the confinement estimates between equation \eqref{eqn: Najmabadi loss rate result} and the finite element code n for electrostatically confined electrons. The legend goes from the top curve to the bottom curve in even steps in the value of $c_0$.}
    \label{fig: error comparison najmabadi}
\end{figure}

In \citet{najmabadi1984collisional}, they determine $c_N = 0.84$, although we have not replicated the calculation used to determine this.
In equation \eqref{eqn: Najmabadi loss rate result}, it is essential to notice that the correction coefficient does change the asymptotic convergence of the problem.
The authors claim that this convergence is enhanced from $\mathcal{O}(1/x_a^2)$ to $\mathcal{O}(\exp(-x_z^2))$, but adding this 0.84 changes where $q_0$ asymptotes to, so it does more than yield faster convergence; it changes the convergence entirely. This is showcased in figure \ref{fig: error comparison najmabadi}, where it is clear that the error asymptotes to a constant level, which depends strongly on $c_N$. Without using this correction factor, the error saturates to around $40\%$ compared to the numerical code, which is quite large and not converging.
With $c_N = 0.84$, equation \ref{eqn: Najmabadi loss rate result} has a roughly 10\% error with the finite element code, whereas \citet{najmabadi1984collisional} reports errors of $\pm 7\%$.
Their expression for $q_0$ is the same as with the LBO, $\ln (w+1/w-1)$, but their $w$ takes the form for large $z_s e \phi / T_s$ of $w = \sqrt{1 + 1/R Z_s}$. Therefore $q_0$ asymptotes at large R to $\ln(1 + 4 R Z_s)$. 
This number is not large and is comparable to the coefficient of $0.84$ they find. 

We do not expect the same coefficient that \citet{najmabadi1984collisional} finds because our approximate distribution function takes a different form, using different powers of velocity. It is worth noting that $w$ asymptotes very differently with the LBO than with the Coulomb operator, which leads to these differences. Since $w = \sqrt{1 + 2 z_s e \phi / (T_s Z_s R)}$ we order $z_s e \phi / T_s < Z_s R $. Then, we get $q_0 = \ln (1 + 2 Z_s R T_s/(z_s e \phi)$, which is also not large and is affected by the convergence factor. The dependence of this $q_0$ on $z_s e \phi / T_s$ affects the convergence, making it more complex to determine the appropriate asymptotic behavior.

Instead, we opt to apply the same numerical technique employed in section \ref{sec: numerics}. By calculating the value of $c_N$ that minimizes the error between our finite element code, we determine the appropriate value of $c_N$. Table \ref{tab: correction factors Najmabadi ions} has the appropriate correction coefficients for a pure hydrogen plasma $(Z_s=1/2)$, and table \ref{tab: correction factors Najmabadi electrons} has the correction coefficients for electrons with a hydrogen background $(Z_s=1)$. Intriguingly, this coefficient depends on the pitch angle scattering rate $Z_s$. It is clear from these tables that \citet{najmabadi1984collisional} calculated this as a constant, but it has a much more complicated dependency in matching the finite element code. It is essential to mention that the collision operator utilized by the finite element code is approximate, especially at low velocities, which impacts the results at low values of $z_s e \phi / T_s$. However, it is intriguing that this coefficient does not asymptote to a constant at large values of $R$ or $z_s e \phi / T_s$, which they suggest should. Furthermore, utilizing this coefficient table effectively provides a much better estimation of the results from the finite element code than the original work. Nonetheless, all figures in the rest of the paper are produced with $c_N=0.84$ as it still is a reasonable estimate.

\begin{table}
\centering
\begin{tabular}{@{}l *{10}{c}@{}}
& \multicolumn{10}{c}{$R$} \\ \cmidrule(lr){2-11}
$z_s e \phi / T_s$ &  5 & 10 & 15 & 20 & 25 & 30 & 35 & 40 & 45 & 50 \\
\midrule
\multicolumn{1}{l|}{1} & 0.876 & 0.860 & 0.777 & 0.770 & 0.691 & 0.702 & 0.680 & 0.653 & 0.649 & 0.679 \\
\multicolumn{1}{l|}{2} & 1.056 & 1.030 & 1.009 & 0.967 & 0.965 & 0.965 & 0.942 & 0.940 & 0.928 & 0.924 \\
\multicolumn{1}{l|}{3} & 1.095 & 1.082 & 1.062 & 1.046 & 1.034 & 1.021 & 1.016 & 1.008 & 0.996 & 1.017 \\
\multicolumn{1}{l|}{4} & 1.114 & 1.107 & 1.091 & 1.078 & 1.067 & 1.058 & 1.051 & 1.044 & 1.039 & 1.035 \\
\multicolumn{1}{l|}{5} & 1.126 & 1.121 & 1.106 & 1.093 & 1.083 & 1.075 & 1.068 & 1.062 & 1.057 & 1.053 \\
\multicolumn{1}{l|}{6} & 1.134 & 1.131 & 1.117 & 1.104 & 1.095 & 1.087 & 1.080 & 1.075 & 1.070 & 1.064 \\
\multicolumn{1}{l|}{7} & 1.136 & 1.132 & 1.117 & 1.104 & 1.095 & 1.086 & 1.080 & 1.073 & 1.068 & 1.066 \\
\multicolumn{1}{l|}{8} & 1.138 & 1.133 & 1.119 & 1.106 & 1.098 & 1.089 & 1.081 & 1.076 & 1.070 & 1.068 \\
\multicolumn{1}{l|}{9} & 1.142 & 1.140 & 1.127 & 1.117 & 1.103 & 1.095 & 1.088 & 1.085 & 1.083 & 1.079 \\
\multicolumn{1}{l|}{10} & 1.142 & 1.141 & 1.123 & 1.118 & 1.106 & 1.093 & 1.084 & 1.079 & 1.079 & 1.091 \\
\bottomrule
\end{tabular}
\caption{A table of optimal correction factors $c_N$ for various values of $z_s e\phi/T_s$ and $R$ to minimize error with the finite element code for hydrogen $Z_s = 1/2$.}
\label{tab: correction factors Najmabadi ions}
\end{table}

\begin{table}
\centering
\begin{tabular}{@{}l *{10}{c}@{}}
& \multicolumn{10}{c}{$R$} \\ \cmidrule(lr){2-11}
$z_s e \phi / T_s$ &  5 & 10 & 15 & 20 & 25 & 30 & 35 & 40 & 45 & 50 \\
\midrule
\multicolumn{1}{l|}{1} & 1.183 & 1.125 & 1.093 & 1.068 & 1.052 & 1.040 & 1.031 & 1.023 & 1.017 & 1.013 \\
\multicolumn{1}{l|}{2} & 1.093 & 1.046 & 1.018 & 0.999 & 0.986 & 0.976 & 0.968 & 0.963 & 0.958 & 0.953 \\
\multicolumn{1}{l|}{3} & 1.081 & 1.042 & 1.018 & 1.002 & 0.991 & 0.983 & 0.976 & 0.971 & 0.966 & 0.963 \\
\multicolumn{1}{l|}{4} & 1.090 & 1.056 & 1.035 & 1.020 & 1.010 & 1.003 & 0.997 & 0.992 & 0.988 & 0.984 \\
\multicolumn{1}{l|}{5} & 1.101 & 1.069 & 1.048 & 1.034 & 1.026 & 1.018 & 1.012 & 1.008 & 1.004 & 1.001 \\
\multicolumn{1}{l|}{6} & 1.112 & 1.083 & 1.062 & 1.048 & 1.041 & 1.033 & 1.028 & 1.023 & 1.020 & 1.014 \\
\multicolumn{1}{l|}{7} & 1.115 & 1.084 & 1.063 & 1.049 & 1.042 & 1.035 & 1.029 & 1.024 & 1.020 & 1.020 \\
\multicolumn{1}{l|}{8} & 1.119 & 1.088 & 1.067 & 1.054 & 1.046 & 1.038 & 1.033 & 1.030 & 1.024 & 1.023 \\
\multicolumn{1}{l|}{9} & 1.127 & 1.097 & 1.078 & 1.066 & 1.054 & 1.050 & 1.043 & 1.042 & 1.041 & 1.035 \\
\multicolumn{1}{l|}{10} & 1.124 & 1.096 & 1.075 & 1.071 & 1.056 & 1.052 & 1.035 & 1.032 & 1.044 & 1.033 \\
\bottomrule
\end{tabular}
\caption{A table of optimal correction factors $c_N$ for various values of $z_s e\phi/T_s$ and $R$ to minimize error with the finite element code for electrons $Z_s = 1$.}
\label{tab: correction factors Najmabadi electrons}
\end{table}
\section{Loss cone boundary conditions}
\label{sec: BC Limits}
From equations \eqref{eqn: BC distribution function}, we have two boundary conditions to fix $q_0$ and $a$. Recall $g(\bar{v},\mu) = \pi^{3/2} \exp(\bar{v}^2) F_s(\bar{v},\mu)$, and hence
\begin{equation}
    g(\rho,z)|_{z=z_0,\rho=0} = 0.
\end{equation}
This is straightforward to apply from equation \eqref{eqn: distg},
\begin{equation}
    q_0 = \left( \ln \left( \frac{w+1}{w-1}\right) \right)^{-1} \label{eqn: q_0 definition appendix},
\end{equation}
where $w = \exp (a^2 - \bar{v}_0^2)$. Equation \eqref{eqn: q_0 definition appendix} can be rewritten as $w = \mathrm{coth}(1/2 q_0)$.

Now, we will expand the second condition in equation \eqref{eqn: BC distribution function}, that the curvature of the contour where our fitted distribution function is zero near the loss cone matches the curvature of the true loss cone. For the sake of generality, we will refer to a variable transformation where the velocity transformation is the same, $z = \exp(\bar{v}^2)$, but the pitch angle transformation is in the general form of $\rho = \bar\rho(\bar{v}) z \sqrt{1-\mu^2}/\mu = \bar\rho(\bar{v}) z \tan \theta \approx  \bar\rho(\bar{v}) z \theta$ where we have isolated the separate pitch angle and velocity dependence. With this form,
\begin{align}
    \partial_{\bar{v}} F(\bar{v},\mu) = \partial_{\bar{v}} \left(\frac{g(\rho,z)}{\pi^{3/2} z}\right)
    =\frac{1}{\pi^{3/2} z} \partial_{\bar{v}} g(\rho,z), 
\end{align}
and
\begin{align}
    \partial_\mu F(\bar{v},\mu) = \partial_\mu \left(\frac{g(\rho,z)}{\pi^{3/2} z} \right) = \frac{1}{\pi^{3/2} z} \partial_\mu g(\rho,z).
\end{align}
So from changing distribution function from $F_s(\bar{v},\mu)$ to $g(\rho,z)$, 
\begin{equation}
    \frac{\partial_{\bar{v}} g}{\partial_\mu g}\bigg\rvert_{z=z_0,\rho=0} = \frac{1}{R \bar{v}_0},
\end{equation}
where we have used $g(0,z_0) = 0$. Continuing, we must expand the partial derivatives in $\bar{v}$ and $\mu$ into their respective derivatives in terms of $\rho$ and $z$ using the chain rule,
\begin{align}
    \partial_{\bar{v}} g(\rho,z) = \frac{\partial g}{\partial \rho} \frac{\partial \rho}{\partial \bar{v}} + \frac{\partial g}{\partial z} \frac{\partial z}{\partial \bar{v}}  
    = \frac{\partial g}{\partial \rho} \rho \left(2 \bar{v}  + \frac{\bar\rho'}{\bar\rho} \right) 
    + \frac{\partial g}{\partial z} 2 \bar{v} z,  \\
    \partial_\mu g = \frac{\partial g}{\partial \rho} \frac{\partial\rho}{\partial \mu} = \frac{\partial g}{\partial \rho} \frac{\partial\rho}{\partial (\cos\theta)} \approx \frac{\partial g}{\partial \rho} \frac{\partial\rho}{-\theta \partial \theta} = \frac{\partial g}{\partial \rho} \frac{\bar\rho(\bar{v}) z }{-\theta} = -\frac{\partial g}{\partial \rho} \frac{\rho}{\theta^2} .
\end{align}
One of the beautiful aspects of this derivation is that this specific choice of variables results in a simple form of the derivative in $\mu$. Furthermore, it can be shown that the derivatives of equation \eqref{eqn: distg} are 
\begin{align}
    \frac{\partial g}{\partial z} \bigg\rvert_{z = z_0, \rho = 0} =& -\frac{2 q_0 w }{z_0 (w^2-1)}, \\
    \frac{\partial g}{\partial \rho} \bigg\rvert_{z = z_0, \rho \ll 1} =& -\frac{q_0 \rho}{2 z_0^2}\left[ \frac{1}{(w+1)^2} -\frac{1}{(w-1)^2}\right].
\end{align}
Now we can plug it all back in together,
\begin{align}
    \frac{1}{R \bar{v}_0} = \frac{\frac{\partial g}{\partial \rho} \rho \left(2 \bar{v}  + \frac{\bar\rho'}{\bar\rho} \right) + \frac{\partial g}{\partial z} 2 \bar{v} z }{ \frac{q_0 }{2 z_0^2}\left[ \frac{1}{(w+1)^2} -\frac{1}{(w-1)^2}\right] \frac{\rho^2}{\theta^2} } \bigg\rvert_{z = z_0, \rho \rightarrow 0}.
\end{align}
Next, we notice that $\rho \rightarrow 0$ and $\partial_\rho g \propto \rho$, so the $\partial_\rho g$ term in the numerator is negligible. We also see that $\rho^2/\theta^2 = \bar \rho^2 z_0^2$ for $z = z_0$ and $\theta \ll 1$, giving
\begin{align}
    \frac{1}{R \bar{v}_0} = \frac{2 \bar{v}_0 }{ \bar\rho(\bar{v}_0)^2}\left( w^2 - 1 \right).
\end{align}
Solving for $w$ we find equation \eqref{eqn: derivative bc matching}. For the collision operator in \citet{najmabadi1984collisional}, $\bar\rho(\bar{v})^2 = 2 \bar{v}^2 / (Z_s - 1/4\bar{v}^2)$, so their equivalent expression would be
\begin{equation}
    w^2 =  1 + \frac{1}{R  (Z_s - 1/4\bar{v}_0^2)} ,
\end{equation}
which agrees with the results of \citet{najmabadi1984collisional}. For Dougherty, we use $\bar\rho(\bar{v})^2 = 4 \bar{v}^4 / Z_s$ so the matching condition becomes the second equality in equation \eqref{eqn: derivative bc matching}.

%% file: main.bbl
\begin{thebibliography}{47}
\expandafter\ifx\csname natexlab\endcsname\relax\def\natexlab#1{#1}\fi
\def\au#1{#1} \def\ed#1{#1} \def\yr#1{#1}\def\at#1{#1}\def\jt#1{\textit{#1}} \def\bt#1{#1}\def\bvol#1{\textbf{#1}} \def\vol#1{#1} \def\pg#1{#1} \def\publ#1{#1}\def\arxiv#1{#1}\def\org#1{#1}\def\st#1{\textit{#1}}

\bibitem[Bagryansky {\em et~al.\/}(2011)Bagryansky, Anikeev, Beklemishev, Donin, Ivanov, Korzhavina, Kovalenko, Kruglyakov, Lizunov, Maximov {\em et~al.\/}]{bagryansky2011confinement}
{\sc \au{Bagryansky, P.A.}, \au{Anikeev, A.V.}, \au{Beklemishev, A.D.}, \au{Donin, A.S.}, \au{Ivanov, A.A.}, \au{Korzhavina, M.S.}, \au{Kovalenko, Yu~V.}, \au{Kruglyakov, E.P.}, \au{Lizunov, A.A.}, \au{Maximov, V.V.} \& \au{others}} \yr{2011}  \at{Confinement of hot ion plasma with $\beta$= 0.6 in the gas dynamic trap}.  \jt{Fusion Science and Technology}  \bvol{59}~(1T),  \pg{31--35}.

\bibitem[Bagryansky {\em et~al.\/}(2015)Bagryansky, Shalashov, Gospodchikov, Lizunov, Maximov, Prikhodko, Soldatkina, Solomakhin \& Yakovlev]{Bagryansky2015}
{\sc \au{Bagryansky, P.~A.}, \au{Shalashov, A.~G.}, \au{Gospodchikov, E.~D.}, \au{Lizunov, A.~A.}, \au{Maximov, V.~V.}, \au{Prikhodko, V.~V.}, \au{Soldatkina, E.~I.}, \au{Solomakhin, A.~L.} \& \au{Yakovlev, D.~V.}} \yr{2015}  \at{Threefold increase of the bulk electron temperature of plasma discharges in a magnetic mirror device}.  \jt{Phys. Rev. Lett.}  \bvol{114},  \pg{205001}.

\bibitem[Beklemishev(2017)]{Beklemishev.2017}
{\sc \au{Beklemishev, A.~D.}} \yr{2017}  \at{{Tail-Waving System for Active Feedback Stabilization of Flute Modes in Open Traps}}.  \jt{Fusion Science and Technology}  \bvol{59}~(1T),  \pg{90--93}.

\bibitem[Beklemishev {\em et~al.\/}(2010)Beklemishev, Bagryansky, Chaschin \& Soldatkina]{beklemishev2010vortex}
{\sc \au{Beklemishev, Alexei~D.}, \au{Bagryansky, Peter~A.}, \au{Chaschin, Maxim~S.} \& \au{Soldatkina, Elena~I.}} \yr{2010}  \at{Vortex confinement of plasmas in symmetric mirror traps}.  \jt{Fusion Science and Technology}  \bvol{57}~(4),  \pg{351--360}.

\bibitem[Berk \& Chen(1988)]{berk1988dissipative}
{\sc \au{Berk, H.L.} \& \au{Chen, C.Y.}} \yr{1988}  \at{Dissipative trapped particle modes in tandem mirrors}.  \jt{The Physics of Fluids}  \bvol{31}~(1),  \pg{137--148}.

\bibitem[Catto \& Bernstein(1981)]{catto1981collisional}
{\sc \au{Catto, Peter~J.} \& \au{Bernstein, Ira~B.}} \yr{1981}  \at{Collisional end losses from conventional and tandem mirrors}.  \jt{The Physics of Fluids}  \bvol{24}~(10),  \pg{1900--1905}.

\bibitem[Catto \& Li(1985)]{catto1985particle}
{\sc \au{Catto, Peter~J.} \& \au{Li, Xing~Zhong}} \yr{1985}  \at{Particle loss rates from electrostatic wells of arbitrary mirror ratios}.  \jt{The Physics of Fluids}  \bvol{28}~(1),  \pg{352--357}.

\bibitem[Celebre {\em et~al.\/}(2023)Celebre, Servidio \& Valentini]{Celebre2023}
{\sc \au{Celebre, G.}, \au{Servidio, S.} \& \au{Valentini, F.}} \yr{2023}  \at{{Phase space dynamics of unmagnetized plasmas: Collisionless and collisional regimes}}.  \jt{Physics of Plasmas}  \bvol{30}~(9),  \pg{092304},  \arxiv{arXiv: https://pubs.aip.org/aip/pop/article-pdf/doi/10.1063/5.0160549/18121857/092304\_1\_5.0160549.pdf}.

\bibitem[Chernin \& Rosenbluth(1978)]{chernin1978ion}
{\sc \au{Chernin, D.P.} \& \au{Rosenbluth, M.N.}} \yr{1978}  \at{Ion losses from end-stoppered mirror trap}.  \jt{Nuclear Fusion}  \bvol{18}~(1),  \pg{47}.

\bibitem[Cohen {\em et~al.\/}(1986)Cohen, Freis \& Newcomb]{cohen1986interchange}
{\sc \au{Cohen, Bruce~I.}, \au{Freis, Robert~P.} \& \au{Newcomb, William~A.}} \yr{1986}  \at{Interchange, rotational, and ballooning stability of long-thin axisymmetric systems with finite-orbit effects}.  \jt{The Physics of Fluids}  \bvol{29}~(5),  \pg{1558--1577}.

\bibitem[Cohen {\em et~al.\/}(1978)Cohen, Rensink, Cutler \& Mirin]{cohen1978collisional}
{\sc \au{Cohen, R.H.}, \au{Rensink, M.E.}, \au{Cutler, T.A.} \& \au{Mirin, A.A.}} \yr{1978}  \at{Collisional loss of electrostatically confined species in a magnetic mirror}.  \jt{Nuclear Fusion}  \bvol{18}~(9),  \pg{1229}.

\bibitem[Dougherty(1964)]{dougherty1964model}
{\sc \au{Dougherty, J.P.}} \yr{1964}  \at{{Model Fokker-Planck equation for a plasma and its solution}}.  \jt{The Physics of Fluids}  \bvol{7}~(11),  \pg{1788--1799}.

\bibitem[Egedal {\em et~al.\/}(2022)Egedal, Endrizzi, Forest \& Fowler]{egedal2022fusion}
{\sc \au{Egedal, J.}, \au{Endrizzi, D.}, \au{Forest, C.B.} \& \au{Fowler, T.K.}} \yr{2022}  \at{Fusion by beam ions in a low collisionality, high mirror ratio magnetic mirror}.  \jt{Nuclear Fusion}  \bvol{62}~(12),  \pg{126053}.

\bibitem[Endrizzi {\em et~al.\/}(2023)Endrizzi, Anderson, Brown, Egedal, Geiger, Harvey, Ialovega, Kirch, Peterson, Petrov {\em et~al.\/}]{endrizzi2023physics}
{\sc \au{Endrizzi, D.}, \au{Anderson, J.K.}, \au{Brown, M.}, \au{Egedal, J.}, \au{Geiger, B.}, \au{Harvey, R.W.}, \au{Ialovega, M.}, \au{Kirch, J.}, \au{Peterson, E.}, \au{Petrov, Yu~V.} \& \au{others}} \yr{2023}  \at{{Physics basis for the Wisconsin HTS Axisymmetric Mirror (WHAM)}}.  \jt{Journal of Plasma Physics}  \bvol{89}~(5),  \pg{975890501}.

\bibitem[Fowler {\em et~al.\/}(2017)Fowler, Moir \& Simonen]{fowler2017new}
{\sc \au{Fowler, T.K.}, \au{Moir, R.W.} \& \au{Simonen, T.C.}} \yr{2017}  \at{A new simpler way to obtain high fusion power gain in tandem mirrors}.  \jt{Nuclear Fusion}  \bvol{57}~(5),  \pg{056014}.

\bibitem[Francisquez {\em et~al.\/}(2020)Francisquez, Bernard, Mandell, Hammett \& Hakim]{francisquez2020conservative}
{\sc \au{Francisquez, Manaure}, \au{Bernard, Tess~N.}, \au{Mandell, Noah~R.}, \au{Hammett, Gregory~W.} \& \au{Hakim, Ammar}} \yr{2020}  \at{{Conservative discontinuous Galerkin scheme of a gyro-averaged Dougherty collision operator}}.  \jt{Nuclear Fusion}  \bvol{60}~(9),  \pg{096021}.

\bibitem[Francisquez {\em et~al.\/}(2022)Francisquez, Juno, Hakim, Hammett \& Ernst]{francisquez2022improved}
{\sc \au{Francisquez, Manaure}, \au{Juno, James}, \au{Hakim, Ammar}, \au{Hammett, Gregory~W} \& \au{Ernst, Darin~R}} \yr{2022}  \at{{Improved multispecies Dougherty collisions}}.  \jt{Journal of Plasma Physics}  \bvol{88}~(3),  \pg{905880303}.

\bibitem[Francisquez {\em et~al.\/}(2023)Francisquez, Rosen, Mandell, Hakim, Forest \& Hammett]{francisquez2023towards}
{\sc \au{Francisquez, Manaure}, \au{Rosen, Maxwell~H.}, \au{Mandell, Noah~R.}, \au{Hakim, Ammar}, \au{Forest, Cary~B.} \& \au{Hammett, Gregory~W.}} \yr{2023}  \at{Toward continuum gyrokinetic study of high-field mirrors}.  \jt{Physics of Plasmas}  \bvol{30}~(10).

\bibitem[Frei {\em et~al.\/}(2022)Frei, Hoffmann \& Ricci]{Frei_Hoffmann_Ricci_2022}
{\sc \au{Frei, B.J.}, \au{Hoffmann, A.C.D.} \& \au{Ricci, P.}} \yr{2022}  \at{Local gyrokinetic collisional theory of the ion-temperature gradient mode}.  \jt{Journal of Plasma Physics}  \bvol{88}~(3),  \pg{905880304}.

\bibitem[Fyfe {\em et~al.\/}(1981)Fyfe, Weiser, Bernstein, Eisenstat \& Schultz]{fyfe1981finite}
{\sc \au{Fyfe, D.}, \au{Weiser, A.}, \au{Bernstein, I.}, \au{Eisenstat, S.} \& \au{Schultz, M.}} \yr{1981}  \at{{A finite element solution of a reduced Fokker-Planck equation}}.  \jt{Journal of Computational Physics}  \bvol{42}~(2),  \pg{327--336}.

\bibitem[Grandgirard {\em et~al.\/}(2016)Grandgirard, Abiteboul, Bigot, Cartier-Michaud, Crouseilles, Dif-Pradalier, Ehrlacher, Esteve, Garbet, Ghendrih, Latu, Mehrenberger, Norscini, Passeron, Rozar, Sarazin, Sonnendrücker, Strugarek \& Zarzoso]{GRANDGIRARD201635}
{\sc \au{Grandgirard, V.}, \au{Abiteboul, J.}, \au{Bigot, J.}, \au{Cartier-Michaud, T.}, \au{Crouseilles, N.}, \au{Dif-Pradalier, G.}, \au{Ehrlacher, Ch.}, \au{Esteve, D.}, \au{Garbet, X.}, \au{Ghendrih, Ph.}, \au{Latu, G.}, \au{Mehrenberger, M.}, \au{Norscini, C.}, \au{Passeron, Ch.}, \au{Rozar, F.}, \au{Sarazin, Y.}, \au{Sonnendrücker, E.}, \au{Strugarek, A.} \& \au{Zarzoso, D.}} \yr{2016}  \at{A 5d gyrokinetic full-f global semi-lagrangian code for flux-driven ion turbulence simulations}.  \jt{Computer Physics Communications}  \bvol{207},  \pg{35--68}.

\bibitem[Hakim {\em et~al.\/}(2020)Hakim, Francisquez, Juno \& Hammett]{hakim2020conservative}
{\sc \au{Hakim, Ammar}, \au{Francisquez, Manaure}, \au{Juno, James} \& \au{Hammett, Gregory~W.}} \yr{2020}  \at{{Conservative discontinuous Galerkin schemes for nonlinear Dougherty--Fokker--Planck collision operators}}.  \jt{Journal of Plasma Physics}  \bvol{86}~(4),  \pg{905860403}.

\bibitem[Hatch {\em et~al.\/}(2013)Hatch, Jenko, Ba\~n\'on Navarro \& Bratanov]{Hatch2013}
{\sc \au{Hatch, D.~R.}, \au{Jenko, F.}, \au{Ba\~n\'on Navarro, A.} \& \au{Bratanov, V.}} \yr{2013}  \at{Transition between saturation regimes of gyrokinetic turbulence}.  \jt{Phys. Rev. Lett.}  \bvol{111},  \pg{175001}.

\bibitem[Hirshman \& Sigmar(1976)]{hirshman1976approximate}
{\sc \au{Hirshman, S.P.} \& \au{Sigmar, D.J.}} \yr{1976}  \at{{Approximate Fokker--Planck collision operator for transport theory applications}}.  \jt{The Physics of Fluids}  \bvol{19}~(10),  \pg{1532--1540}.

\bibitem[Hoffmann {\em et~al.\/}(2023)Hoffmann, Frei \& Ricci]{Hoffmann_Frei_Ricci_2023}
{\sc \au{Hoffmann, A.C.D.}, \au{Frei, B.J.} \& \au{Ricci, P.}} \yr{2023}  \at{Gyrokinetic simulations of plasma turbulence in a z-pinch using a moment-based approach and advanced collision operators}.  \jt{Journal of Plasma Physics}  \bvol{89}~(2),  \pg{905890214}.

\bibitem[Jackson(1999)]{jackson1999classical}
{\sc \au{Jackson, John~David}} \yr{1999} Classical electrodynamics.

\bibitem[Khudik(1997)]{khudik1997longitudinal}
{\sc \au{Khudik, V.N.}} \yr{1997}  \at{Longitudinal losses of electrostatically confined particles from a mirror device with arbitrary mirror ratio}.  \jt{Nuclear Fusion}  \bvol{37}.

\bibitem[Knyazev {\em et~al.\/}(2023)Knyazev, Dorf \& Krasheninnikov]{Knyazev2023}
{\sc \au{Knyazev, A.R.}, \au{Dorf, M.} \& \au{Krasheninnikov, S.I.}} \yr{2023}  \at{Implementation and verification of a model linearized multi-species collision operator in the cogent code}.  \jt{Computer Physics Communications}  \bvol{291},  \pg{108829}.

\bibitem[Lenard \& Bernstein(1958)]{lenard1958plasma}
{\sc \au{Lenard, Andrew} \& \au{Bernstein, Ira~B.}} \yr{1958}  \at{Plasma oscillations with diffusion in velocity space}.  \jt{Physical Review}  \bvol{112}~(5),  \pg{1456}.

\bibitem[Loureiro {\em et~al.\/}(2016)Loureiro, Dorland, Fazendeiro, Kanekar, Mallet, Vilelas \& Zocco]{LOUREIRO2016}
{\sc \au{Loureiro, N.F.}, \au{Dorland, W.}, \au{Fazendeiro, L.}, \au{Kanekar, A.}, \au{Mallet, A.}, \au{Vilelas, M.S.} \& \au{Zocco, A.}} \yr{2016}  \at{Viriato: A fourier–hermite spectral code for strongly magnetized fluid–kinetic plasma dynamics}.  \jt{Computer Physics Communications}  \bvol{206},  \pg{45--63}.

\bibitem[Majeski {\em et~al.\/}(2017)Majeski, Bell, Boyle, Kaita, Kozub, LeBlanc, Lucia, Maingi, Merino, Raitses, Schmitt, Allain, Bedoya, Bialek, Biewer, Canik, Buzi, Koel, Patino, Capece, Hansen, Jarboe, Kubota, Peebles \& Tritz]{Majeski2017}
{\sc \au{Majeski, R.}, \au{Bell, R.~E.}, \au{Boyle, D.~P.}, \au{Kaita, R.}, \au{Kozub, T.}, \au{LeBlanc, B.~P.}, \au{Lucia, M.}, \au{Maingi, R.}, \au{Merino, E.}, \au{Raitses, Y.}, \au{Schmitt, J.~C.}, \au{Allain, J.~P.}, \au{Bedoya, F.}, \au{Bialek, J.}, \au{Biewer, T.~M.}, \au{Canik, J.~M.}, \au{Buzi, L.}, \au{Koel, B.~E.}, \au{Patino, M.~I.}, \au{Capece, A.~M.}, \au{Hansen, C.}, \au{Jarboe, T.}, \au{Kubota, S.}, \au{Peebles, W.~A.} \& \au{Tritz, K.}} \yr{2017}  \at{{Compatibility of lithium plasma-facing surfaces with high edge temperatures in the Lithium Tokamak Experiment}}.  \jt{Physics of Plasmas}  \bvol{24}~(5),  \pg{056110},  \arxiv{arXiv: https://pubs.aip.org/aip/pop/article-pdf/doi/10.1063/1.4977916/15997017/056110\_1\_online.pdf}.

\bibitem[Mandell {\em et~al.\/}(2022)Mandell, Dorland, Abel, Gaur, Kim, Martin \& Qian]{mandell2022gx}
{\sc \au{Mandell, N.R.}, \au{Dorland, W.}, \au{Abel, I.}, \au{Gaur, R.}, \au{Kim, P.}, \au{Martin, M.} \& \au{Qian, T.}} \yr{2022}  \at{{GX: a GPU-native gyrokinetic turbulence code for tokamak and stellarator design}}.  \jt{arXiv preprint arXiv:2209.06731} .

\bibitem[Najmabadi {\em et~al.\/}(1984)Najmabadi, Conn \& Cohen]{najmabadi1984collisional}
{\sc \au{Najmabadi, Farrokh}, \au{Conn, R.W.} \& \au{Cohen, Ronald~H.}} \yr{1984}  \at{Collisional end loss of electrostatically confined particles in a magnetic mirror field}.  \jt{Nuclear Fusion}  \bvol{24}~(1),  \pg{75}.

\bibitem[Ochs {\em et~al.\/}(2023)Ochs, Munirov \& Fisch]{ochs2023confinement}
{\sc \au{Ochs, Ian~E.}, \au{Munirov, Vadim~R.} \& \au{Fisch, Nathaniel~J.}} \yr{2023}  \at{Confinement time and ambipolar potential in a relativistic mirror-confined plasma}.  \jt{Physics of Plasmas}  \bvol{30}~(5).

\bibitem[Parker \& Dellar(2015)]{Parker_Dellar_2015}
{\sc \au{Parker, Joseph~T.} \& \au{Dellar, Paul~J.}} \yr{2015}  \at{Fourier–hermite spectral representation for the vlasov–poisson system in the weakly collisional limit}.  \jt{Journal of Plasma Physics}  \bvol{81}~(2),  \pg{305810203}.

\bibitem[Pastukhov(1974)]{pastukhov1974collisional}
{\sc \au{Pastukhov, V.P.}} \yr{1974}  \at{Collisional losses of electrons from an adiabatic trap in a plasma with a positive potential}.  \jt{Nuclear Fusion}  \bvol{14}~(1),  \pg{3}.

\bibitem[Perrone {\em et~al.\/}(2020)Perrone, Jorge \& Ricci]{Perone2020}
{\sc \au{Perrone, L.~M.}, \au{Jorge, R.} \& \au{Ricci, P.}} \yr{2020}  \at{{Four-dimensional drift-kinetic model for scrape-off layer plasmas}}.  \jt{Physics of Plasmas}  \bvol{27}~(11),  \pg{112502},  \arxiv{arXiv: https://pubs.aip.org/aip/pop/article-pdf/doi/10.1063/5.0024968/16099478/112502\_1\_online.pdf}.

\bibitem[Pezzi {\em et~al.\/}(2016)Pezzi, Camporeale \& Valentini]{Pezzi2016}
{\sc \au{Pezzi, Oreste}, \au{Camporeale, Enrico} \& \au{Valentini, Francesco}} \yr{2016}  \at{{Collisional effects on the numerical recurrence in Vlasov-Poisson simulations}}.  \jt{Physics of Plasmas}  \bvol{23}~(2),  \pg{022103},  \arxiv{arXiv: https://pubs.aip.org/aip/pop/article-pdf/doi/10.1063/1.4940963/15794445/022103\_1\_online.pdf}.

\bibitem[Post \& Rosenbluth(1966)]{post1966electrostatic}
{\sc \au{Post, Richard~F.} \& \au{Rosenbluth, M.N.}} \yr{1966}  \at{Electrostatic instabilities in finite mirror-confined plasmas}.  \jt{The Physics of Fluids}  \bvol{9}~(4),  \pg{730--749}.

\bibitem[Rosenbluth {\em et~al.\/}(1957)Rosenbluth, MacDonald \& Judd]{rosenbluth1957fokker}
{\sc \au{Rosenbluth, Marshall~N.}, \au{MacDonald, William~M.} \& \au{Judd, David~L.}} \yr{1957}  \at{{Fokker-Planck equation for an inverse-square force}}.  \jt{Physical Review}  \bvol{107}~(1),  \pg{1}.

\bibitem[Ryutov {\em et~al.\/}(2011)Ryutov, Berk, Cohen, Molvik \& Simonen]{ryutov2011magneto}
{\sc \au{Ryutov, D.D.}, \au{Berk, H.L.}, \au{Cohen, B.I.}, \au{Molvik, A.W.} \& \au{Simonen, T.C.}} \yr{2011}  \at{Magneto-hydrodynamically stable axisymmetric mirrors}.  \jt{Physics of Plasmas}  \bvol{18}~(9).

\bibitem[Sharma \& Hammett(2011)]{sharma2011fast}
{\sc \au{Sharma, Prateek} \& \au{Hammett, Gregory~W}} \yr{2011}  \at{A fast semi-implicit method for anisotropic diffusion}.  \jt{Journal of Computational Physics}  \bvol{230}~(12),  \pg{4899--4909}.

\bibitem[Taitano {\em et~al.\/}(2015)Taitano, Chac{\'o}n, Simakov \& Molvig]{taitano2015mass}
{\sc \au{Taitano, William~T.}, \au{Chac{\'o}n, Luis}, \au{Simakov, A.N.} \& \au{Molvig, K.}} \yr{2015}  \at{{A mass, momentum, and energy conserving, fully implicit, scalable algorithm for the multi-dimensional, multi-species Rosenbluth--Fokker--Planck equation}}.  \jt{Journal of Computational Physics}  \bvol{297},  \pg{357--380}.

\bibitem[Ulbl {\em et~al.\/}(2023)Ulbl, Body, Zholobenko, Stegmeir, Pfennig \& Jenko]{ulbl2023influence}
{\sc \au{Ulbl, Philipp}, \au{Body, Thomas}, \au{Zholobenko, Wladimir}, \au{Stegmeir, Andreas}, \au{Pfennig, Jan} \& \au{Jenko, Frank}} \yr{2023}  \at{Influence of collisions on the validation of global gyrokinetic simulations in the edge and scrape-off layer of tcv}.  \jt{Physics of Plasmas}  \bvol{30}~(5).

\bibitem[Ulbl {\em et~al.\/}(2022)Ulbl, Michels \& Jenko]{ulbl2022implementation}
{\sc \au{Ulbl, Philipp}, \au{Michels, Dominik} \& \au{Jenko, Frank}} \yr{2022}  \at{{Implementation and verification of a conservative, multi-species, gyro-averaged, full-f, Lenard-Bernstein/Dougherty collision operator in the gyrokinetic code GENE-X}}.  \jt{Contributions to Plasma Physics}  \bvol{62}~(5-6),  \pg{e202100180},  \arxiv{arXiv: https://onlinelibrary.wiley.com/doi/pdf/10.1002/ctpp.202100180}.

\bibitem[White {\em et~al.\/}(2018)White, Hassam \& Brizard]{white2018centrifugal}
{\sc \au{White, Roscoe}, \au{Hassam, Adil} \& \au{Brizard, Alain}} \yr{2018}  \at{Centrifugal particle confinement in mirror geometry}.  \jt{Physics of Plasmas}  \bvol{25}~(1).

\bibitem[Ye {\em et~al.\/}(2024)Ye, Hu, Shu \& Zhong]{Ye2024}
{\sc \au{Ye, Boyang}, \au{Hu, Jingwei}, \au{Shu, Chi-Wang} \& \au{Zhong, Xinghui}} \yr{2024}  \at{Energy-conserving discontinuous galerkin methods for the vlasov-ampère system with dougherty-fokker-planck collision operator}.  \jt{Journal of Computational Physics}  \bvol{514},  \pg{113219}.

\end{thebibliography}
